\documentclass[preprint2]{aastex}

\shorttitle{Galaxy Stellar Mass Assembly between $0.2<z<2$}
\shortauthors{}
 
\begin{document}


\title{Galaxy Stellar Mass Assembly between $0.2<z<2$ \\ from the S-COSMOS survey \altaffilmark{1}}


\author{
O. Ilbert\altaffilmark{2,3}, 
M. Salvato\altaffilmark{4}, 
E. Le Floc'h\altaffilmark{2}, 
H. Aussel\altaffilmark{5},
P. Capak\altaffilmark{4,6}, 
H. J. McCracken\altaffilmark{7},
B. Mobasher\altaffilmark{8}, 
J. Kartaltepe\altaffilmark{2},
N. Scoville\altaffilmark{4}, 
D. B. Sanders\altaffilmark{2},
S. Arnouts\altaffilmark{9}, 
K. Bundy\altaffilmark{10},
P. Cassata\altaffilmark{3}, 
J.-P. Kneib\altaffilmark{3}, 
A. Koekemoer\altaffilmark{11}, 
O. Le F\`evre\altaffilmark{3}, 
S. Lilly\altaffilmark{12},
J. Surace\altaffilmark{6}, 
Y. Taniguchi\altaffilmark{13},
L. Tasca\altaffilmark{10}, 
D. Thompson\altaffilmark{4,14},
L. Tresse\altaffilmark{3}, 
M. Zamojski\altaffilmark{4},
G. Zamorani\altaffilmark{15}, 
E. Zucca\altaffilmark{15}
\email{}
}



\begin{abstract}
We follow the galaxy stellar mass assembly by morphological and
spectral type in the COSMOS 2-deg$^2$ field. We derive the stellar
mass functions and stellar mass densities from $z=2$ to $z=0.2$ using
196,000 galaxies selected at $F_{3.6\mu m}> 1 \mu Jy$ with accurate
photometric redshifts
($\sigma_{(z_{phot}-z_{spec})/(1+z_{spec})}=0.008$ at $i^+<22.5$).
Using a spectral classification, we find that $z\sim 1$ is an epoch of
transition in the stellar mass assembly of quiescent galaxies. Their
stellar mass density increases by 1.1 dex between $z=1.5-2$ and
$z=0.8-1$ ($\Delta t\sim 2.5$ Gyr), but only by 0.3 dex between
$z=0.8-1$ and $z \sim 0.1$ ($\Delta t\sim 6$ Gyr). Then, we add the
morphological information and find that 80-90\% of the massive
quiescent galaxies ($log({\cal M}) \sim 11$) have an elliptical
morphology at $z<0.8$. Therefore, a dominant mechanism links the
shutdown of star formation and the acquisition of an elliptical
morphology in massive galaxies.  Still, a significant fraction of
quiescent galaxies present a Spi/Irr morphology at low mass (40-60\%
at $log({\cal M})\sim 9.5$), but this fraction is smaller than
predicted by semi-analytical models using a ``halo quenching''
recipe. We also analyze the evolution of star-forming galaxies and
split them into ``intermediate activity'' and ``high activity''
galaxies. We find that the most massive ``high activity'' galaxies end
their high star formation rate phase first. Finally, the space density
of massive star-forming galaxies becomes lower than the space density
of massive elliptical galaxies at $z<1$. As a consequence, the rate of
``wet mergers'' involved in the formation of the most massive
ellipticals must decline very rapidly at $z<1$, which could explain
the observed slow down in the assembly of these quiescent and massive
sources.
\end{abstract}





\altaffiltext{1}{Based on observations with the NASA/ESA {\em
Hubble Space Telescope}, obtained at the Space Telescope Science
Institute, which is operated by AURA Inc, under NASA contract NAS
5-26555. Also based on observations made with the Spitzer Space
Telescope, which is operated by the Jet Propulsion Laboratory,
California Institute of Technology, under NASA contract 1407. Also
based on data collected at : the Subaru Telescope, which is operated
by the National Astronomical Observatory of Japan; the XMM-Newton, an
ESA science mission with instruments and contributions directly funded
by ESA Member States and NASA; the European Southern Observatory under
Large Program 175.A-0839, Chile; Kitt Peak National Observatory, Cerro
Tololo Inter-American Observatory and the National Optical Astronomy
Observatory, which are operated by the Association of Universities for
Research in Astronomy, Inc.  (AURA) under cooperative agreement with
the National Science Foundation; and the Canada-France-Hawaii
Telescope with MegaPrime/MegaCam operated as a joint project by the
CFHT Corporation, CEA/DAPNIA, the NRC and CADC of Canada, the CNRS of
France, TERAPIX and the Univ. of Hawaii.}

\altaffiltext{2}{Institute for Astronomy, 2680 Woodlawn Dr., University of Hawaii, Honolulu, Hawaii, 96822}
\altaffiltext{3}{Laboratoire d'Astrophysique de Marseille, Universit\'e de Provence, CNRS, BP 8, Traverse du Siphon, 13376 Marseille Cedex 12, France}
\altaffiltext{4}{California Institute of Technology, MC 105-24, 1200 East California Boulevard, Pasadena, CA 91125}
\altaffiltext{5}{AIM Unit\'e Mixte de Recherche CEA  CNRS Universit\'e Paris VII UMR n158}
\altaffiltext{6}{Spitzer Science Center, California Institute of Technology, Pasadena, CA 91125}
\altaffiltext{7}{Institut d'Astrophysique de Paris, UMR7095 CNRS, Universit\'e Pierre et Marie Curie, 98 bis Boulevard Arago, 75014 Paris, France}
\altaffiltext{8}{Department of Physics and Astronomy, University of California, Riverside, CA, 92521, USA}
\altaffiltext{9}{Canada France Hawaii telescope corporation, 65-1238 Mamalahoa Hwy, Kamuela, Hawaii 96743, USA}
\altaffiltext{10}{Department of Astronomy and Astrophysics, University of Toronto, 50 St. George Street, Room 101, Toronto, ON M58 3H4, Canada}
\altaffiltext{11}{Space Telescope Science Institute, 3700 San Martin Drive, Baltimore, MD 21218}
\altaffiltext{12}{Department of Physics, ETH Zurich, CH-8093 Zurich, Switzerland}
\altaffiltext{13}{Research Center for Space and Cosmic Evolution, Ehime University, 
        Bunkyo-cho 2-5, Matsuyama 790-8577, Japan}
\altaffiltext{14}{LBT Observatory, University of Arizona, 933 N. Cherry Ave., Tucson, Arizona, 85721-0065, USA}
\altaffiltext{15}{INAF-Osservatorio Astronomico di Bologna, via Ranzani 1, I-40127 Bologna, Italy}

\section{Introduction}\label{Introduction}

A clear and comprehensive picture describing the physical processes
which regulate stellar mass growth in galaxies is still missing in our
understanding of galaxy evolution. Indeed, the stellar mass growth is
regulated by a complex interplay between the radiative cooling of the
gas (e.g. White 1978), cold accretion (e.g. Kere{\v s} et al. 2005),
the spatial redistribution of the gas along the hierarchical growth of
dark matter halos (e.g. Springel et al. 2006) and the feedback from
supernovae and Active Galaxy Nuclei (e.g. Benson et al. 2003, Croton
et al. 2006). AGN feedback is a central process recently added to
galaxy formation models in order to suppress excessive cooling of the
gas in massive halos (e.g. Bower et al. 2006, Croton et al. 2006,
Menci et al. 2006, Cattaneo et al. 2006).  Even with the inclusion of
AGN feedback, semi-analytical models still miss a population of
massive galaxies at $z\sim 2$ (McCracken et al. 2009) and
overproduce the number density of low mass galaxies (e.g. Kitzbichler
\& White 2008, Stringer et al. 2008). Therefore, a better description
of star formation activity is still needed. The stellar mass function
(MF), as studied in this paper, characterizes how star formation
activity build the stellar mass of each galaxy type.

Merging between galaxies is another central mechanism in stellar mass
assembly. However, there appears to be little consensus between direct
estimates of the merger rate (e.g. Le F\`evre et al. 2000, Kartaltepe
et al. 2007, Lotz et al. 2008). An alternative approach is to study
the product of major mergers. Indeed, these are expected to deplete
the low mass end of the MF in favor of high-mass galaxies, and to
produce galaxies with elliptical morphologies (e.g. Toomre \& Toomre
1972). Therefore, a detailed measurement of the MF by galaxy type can
yield valuable clues on galaxy assembly by mergers. This measurement
can also be considered to be a crucial test of the hierarchical
paradigm since the assembly of elliptical galaxies is expected to
follow a hierarchical build-up similar to that of their host dark
matter halos (e.g. Kauffmann et al. 1993, de Lucia et al. 2006).

Following the stellar mass assembly of a given galaxy population
requires that the sample be split into well characterized galaxy
types. A multi-color classification scheme is often the only possible
method to split the faint high redshift samples by type. The bimodal
distribution of the galaxies in a color - magnitude diagram is a
common tool often used to differentiate two populations: ``blue
cloud'' and ``red sequence'' galaxies (e.g. Bell et al. 2004, Faber et
al. 2007, Franzetti et al. 2007).  The red sequence galaxies include
mostly passive galaxies with an elliptical morphology (e.g. Strateva
et al. 2001, Cassata et al. 2007), but also a significant fraction of
dust extincted star-forming galaxies (e.g. William et al. 2008) and
Spi/Irr galaxies with a quenched star-formation (e.g. Bell et
al. 2008). A novel color-color selection technique ($M_U-M_V$ versus
$M_V-M_J$) has been proposed by William et al. (2008). This
color-color selection breaks the degeneracies between dust-extincted
star-forming galaxies and those with quenched star-formation. This
diagram is more efficient for detecting a bimodal distribution than a
color-magnitude plot (William et al. 2008). An alternative multi-color
classification method is based on a template-fitting procedure
(e.g. Lin et al. 1999, Wolf et al. 2003, Zucca et al. 2006). The
advantage of this method is that it defines more than two spectral
types. But the different template selections are difficult to compare
from one study to another.

However, the spectral classifications are sensitive to the
instantaneous star formation rate (SFR). Different galaxy populations
mixed in the same spectral class can be disentangled by adding
morphological information. Automatic morphological classifications
(e.g. Abraham et al. 1996) performed on high resolution images are
efficient for discriminating at least two robust classes: E/S0 and
Spi/Irr galaxies (e.g. Lauger et al. 2005, Menanteau et al. 2006, Lotz
et al. 2008, Capak et al. 2008).  The combination of morphological and
spectral classifications allow us to isolate the ``blue elliptical''
galaxies (e.g. Cross et al. 2004, Menanteau et al. 2006, Ilbert et
al. 2006a) which could include newly-formed ellipticals still
harboring star-formation (e.g. Van Dokkum \& Franx 2001), Spi/Irr
with quenched star formation (Bell et al. 2008) and passive elliptical
galaxies (e.g. Abraham et al. 2007).

Stellar mass assembly in galaxies by spectral and morphological type
has already been investigated using deep optical and near-infrared
(NIR) surveys. Bundy et al. (2005), Franceschini et al. (2006) and
Pannella et al. (2006) have derived the MF by morphological type using
respectively 2150, 1478 and 1645 galaxies at $z<1.4$ in the two GOODS
fields covering 160 arcmin$^2$ each (Giavalisco et al. 2004). Borch et
al. (2006) and Bundy et al. (2006) derived the MF for blue cloud and
red sequence galaxies using larger fields of 0.8 deg$^2$ and 1.5
deg$^2$, respectively.  These analyzes showed that massive elliptical
or red sequence galaxies are already in place at $z\sim 1$, while the
density is still increasing at lower masses. Vergani et al. confirmed
these results using the 4000${\rm \AA}$ Balmer break to separate
galaxy populations in early and late type systems. Therefore, the
``downsizing'' pattern found by Cowie et al. (1996) could be extended
to the assembly process of ellipticals at $z<1$ (e.g. Cimatti et
al. 2006).  Using the K-band luminosity function rather than the MF,
Arnouts et al. (2008) and Cirasuolo et al. (2007) were able to study
the stellar mass assembly for red sequence and blue cloud galaxies at
$z>1$. They found a rapid rise in the space density of massive red
sequence galaxies from $z \sim 2$ to $z \sim 1$ (Cirasuolo et al. 2007
and Arnouts et al. 2007).  Abraham et al. (2007) combined morphology
and colors to study stellar mass evolution for 144 galaxies with
spectroscopic redshifts at $z>0.8$ using data from the {\it Gemini
Deep Deep Survey}. They confirmed the importance of this redshift
range in the birth of passive elliptical galaxies.

This paper presents the evolution of the galaxy stellar mass function
and stellar mass density using the COSMOS survey.  This survey
(Scoville et al. 2007) provides four main advantages over previous
studies that have attempted to measure MF evolution: {\it 1)} it
covers 2-deg$^2$ which reduces the effect of cosmic variance; {\it 2)}
a morphological classification can be carried out based on the {\it
Hubble Space Telescope}-Advanced Camera for Surveys ({\it HST}/ACS)
images (Koekemoer et al. 2007); {\it 3)} deep {\it Spitzer}/IRAC
($3.6-8.0 \mu m$) (Sanders et al. 2007) and CFHT/WIRCAM $K_{\rm
s}$-band data (McCracken et al. 2009)  allow us to estimate
accurate stellar masses out to $z \sim 2$; {\it 4)} the extensive
multi-$\lambda$ coverage of COSMOS provides accurate photometric
redshifts (Ilbert et al. 2009) that can be used to derive the galaxy
stellar MF. We took special care to characterize the galaxy
populations, including galaxy morphologies. A first study by Scarlata
et al. (2007) in the COSMOS field already combined morphological and
spectral classifications to study the B-band luminosity function. We
supplement this study by deriving the stellar MF. We provide an
estimate of the MF which simultaneously covers a large range of
redshift ($0.2<z<2$) and a large range of stellar masses ($10^9< {\cal
M} / {\cal M_{\Sun}}<10^{12}$) using K-band images that are 1.5 mag
deeper than those used by Bundy et al. (2006). We also combine
morphological and spectral classifications over a field 20$\times$ and
100$\times$ larger than Bundy et al. (2005) and Abraham et al. (2007),
respectively.

The COSMOS data are introduced in \S\ref{Data}. The criteria used to
split the galaxy sample into various populations are described in
\S\ref{classification}. We present the method used to compute the
galaxy stellar masses in \S\ref{method}. \S\ref{MFtot},
\S\ref{MFearly}, \S\ref{late} and \S\ref{blueEll} present the stellar
MF and stellar mass density of total, early and late type samples,
respectively. The results are discussed in
\S\ref{discussion}. Throughout this paper, we use the standard
cosmology ($\Omega_m~=~0.3$, $\Omega_\Lambda~=~0.7$) with
$H_{\rm0}~=~70$~km~s$^{-1}$~Mpc$^{-1}$). Magnitudes are given in the
$AB$ system. The stellar masses are given in units of solar masses
(${\cal M}_\Sun$) for a Chabrier initial mass function (hereafter
IMF). The stellar masses based on a Salpeter IMF (Arnouts et
al. 2007), ``diet'' Salpeter IMF (Bell et al. 2008) and Kroupa IMF
(Borch et al. 2006) were converted into a Chabrier IMF by adding -0.24
dex, -0.09 dex, and 0 dex, respectively, to the logarithm of the
stellar masses.

\section{Data}\label{Data}

\subsection{The 3.6$\mu m$ selected catalogue}\label{IRACcatalogue}

This analysis is based on a mass selected sample as generated from the
$3.6\mu m$ IRAC catalogue of the S-COSMOS survey (Sanders et
al. 2007).

The IRAC data were taken during the Spitzer Cycle 2 S-COSMOS survey,
which used 166 hrs to map the full 2-deg$^2$ COSMOS field (centered at
J2000 RA =10:00:28.6, Dec = +02:12:21.0). The observations were
carried out in 4 channels: 3.6$\mu$m, 4.5$\mu$m, 5.6$\mu$m and
8.0$\mu$m. The data were initially processed by the Spitzer Science
Center (SSC). The raw scientific exposures were flux calibrated and
corrected for well-understood instrumental signatures using a pipeline
described by Surace et al. (2005).  Once the frame-level images were
prepared, they were projected onto a common tangent projection and
coadded using the SSC MOPEX
software\footnote{http://ssc.spitzer.caltech.edu/postbcd/}. The images
and the corresponding uncertainty maps were generated for each of the
4 channels.

\begin{figure}[htb!]
\includegraphics[width=7.9cm]{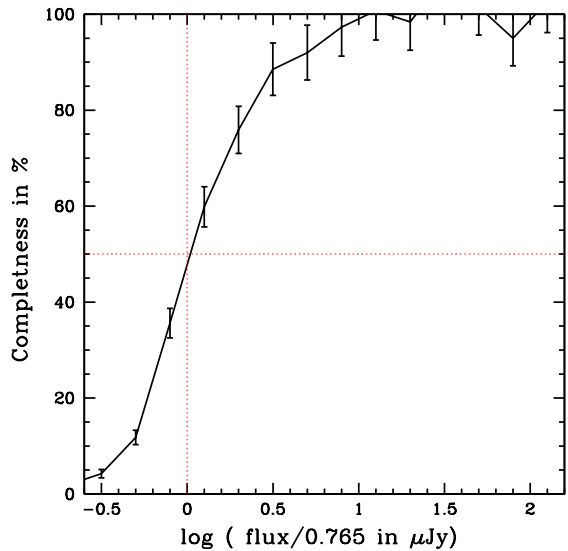}
\caption{Completeness at 3.6$\mu m$: fraction of sources simulated
  in the 3.6$\mu m$ image which are detected with SExtractor, as a
  function of flux. \label{completeness}}
\end{figure}

\begin{figure}[htb!]
\includegraphics[width=7.9cm]{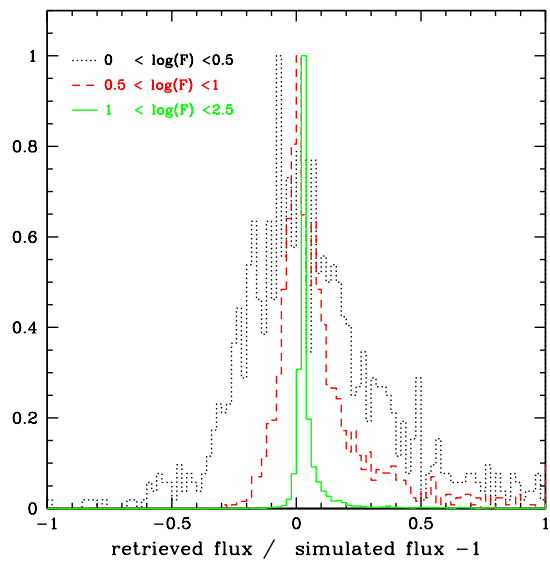}
\caption{Ratio between the 3.6$\mu m$ flux measured with SExtractor
  over the simulated flux. \label{IRAC_flux}}
\end{figure}

The source catalogue was extracted using the SExtractor software
(Bertin \& Arnouts, 1996). The source detection is performed at
3.6$\mu m$. The IRAC 3.6$\mu m$ images have a Point Spread Function
(PSF) of 1.7$^{\prime\prime}$ which necessitates a careful deblending
of the sources. This was obtained with a Mexican Hat filtering of the
images by SExtractor. In order to estimate the completeness of the
3.6$\mu $m catalogue, we simulated point-like sources in the 3.6$\mu
$m mosaic. We simulated simultaneously 10,000 sources with a flux
ranging from 0.1 to 300 $\mu Jy$. The simulated sources were
distributed randomly in the field without any a priory knowledge of
the position of the real 3.6$\mu $m sources (these sources can fall
behind or nearby a real bright source). We run SExtractor on this new
image using exactly the same configuration as for real data. Finally,
we estimated the fraction of simulated sources that we are able to
detect, as a function of flux (see Figure \ref{completeness}). We
found that the IRAC catalogue is 90\% complete at 5 $\mu Jy$ and 50\%
complete at 1 $\mu Jy$.

We also used the SExtractor software the measure the IRAC fluxes.
Following Surace et al. (2005), the fluxes were measured over a
circular aperture of radius 1.9$^{\prime\prime}$. This small radius
provides a flux measurement less affected by the presence of nearby
sources. We tested the accuracy of the fluxes recovered by SExtractor
using the simulation described previously.  Figure \ref{IRAC_flux}
shows the comparison between the simulated flux and the flux measured
by SExtractor. We obtained a flux accuracy of 5\%, 10\% and 25\% for
sources at $10 < F < 300 \mu Jy$, $3< F < 10 \mu Jy$ and $1< F < 3 \mu
Jy$, respectively. These uncertainties on the 3.6$\mu m$ fluxes are
not directly propagated into the stellar masses since deep
near-infrared data ($J$, $H$, $K$) as well as $24$ optical bands
constrain the rescaling of the best-fit fit templates. Based on the
same simulation, we derived an aperture correction of 1.31 at 3.6$\mu
m$ to convert the aperture flux to total flux (assuming the sources to
be point-like). However, we caution the reader that specific software
like CONVPHOT (De Santis et al. 2006) or TFIT (Laidler et al. 2007)
could provide flux measurements less affected by the confusion by
using the K band image as a prior.

Finally, we masked the brightest sources ($K_{\rm s}<12$), as well as
poor image quality areas and the field boundaries. After removing the
masked areas, the $3.6\mu m$ catalogue contains a total of 306,000
sources brighter than $1 \mu Jy$ (50\% completness limit) over an area
of 2.3-$deg^2$.

\subsection{Optical and photo-z catalogues}\label{photo-z_catalogue}

We cross-matched the 3.6$\mu m$ catalogue with the COSMOS photometric
(Capak et al. 2008) and photo-z catalogs (Ilbert et al. 2009).  The
photo-z were derived for all of the sources in the COSMOS photometric
catalogue (1,500,515 sources in total, 937,013 sources at
$i^+<26.5$). The photometric fluxes are measured in 31 bands (2 bands
from GALEX, 6 broad bands from SuprimeCam/Subaru camera, 2 broad bands
from MEGACAM at CFHT, 14 medium and narrow bands from
SuprimeCam/Subaru, $J$-band from the WFCAM/UKIRT camera, $H$ and
$K$-band from the WIRCAM/CFHT camera, and the 4 IRAC/Spitzer
channels). The imaging data are extremely deep, reaching $u^* \sim
27$, $i^+ \sim 26.2$ and $K_{\rm s}\sim 23.7$ for a $5\sigma$
detection in a 3$^{\prime\prime}$ aperture (the sensitivities are
listed in Capak et al. 2008 and Salvato et al. 2009). We restricted
this study to the area covered by the deep optical Subaru image
(2-deg$^2$, $149.4114<\alpha<150.8269$ and $1.4987<\delta<2.9127$) in
order to assure a robust photo-z estimate.

We derived photometric redshifts using the {\it Le
  Phare}\footnote{www.cfht.hawaii.edu/~arnouts/LEPHARE/cfht\_lephare} code
(Arnouts et al. 2002 and Ilbert et al. 2006b) with a $\chi^2$
template-fitting method. The photo-z have been updated in
  comparison to Ilbert et al. (2009) by including new H band data. The
  photo-z are estimated using the median of the Probability
  Distribution Function (PDFz) rather than the minimum of the $\chi^2$
  distribution. The photo-z were calibrated with 4,148 spectroscopic
  redshifts at $i^+_{AB}<22.5$ from the zCOSMOS survey (Lilly et
  al. 2009, in prep.). The comparison between the photometric and
  spectroscopic redshifts shows that the fraction of outliers (defined
  as galaxies with $(z_{phot}-z_{spec})/(1+z_{spec}) >0.15$) is less
  than 1\% and the accuracy is as good as
  $\sigma_{(z_{phot}-z_{spec})/(1+z_{spec})}=0.008$ at $i^+_{AB}
  <22.5$. A spectroscopic follow-up of $24 \mu m$ selected sources
  at $z<1.5$ (Kartaltepe et al. 2009, in prep.) allows us to
  characterize the photo-z accuracy at fainter magnitude. We found
  $\sigma_{(z_{phot}-z_{spec})/(1+z_{spec})}=0.011$ at $22.5<i^+_{AB}
  <24$ and $\sigma_{(z_{phot}-z_{spec})/(1+z_{spec})}=0.053$ at
  $24<i^+_{AB} <25$ for this infrared selected sample. At $z>1.5$, we
  used the zCOSMOS-faint spectroscopic sample (Lilly et al. 2009, in
  prep.) to quantify the quality of the photo-z in the
  magnitude/redshift range where the photo-z are expected to have the
  highest uncertainty. These color selected galaxies have median
  apparent magnitude of $i_{\rm med}^+\sim 24.1$ and a median redshift
  of 2.1. At $1.5<z<3$, we obtained an accuracy of $\sigma_{\Delta
    z/(1+z)}=0.04$ with 10\% of catastrophic failures. However, these
  various spec-z samples probe only specific populations (infrared
  selected, color selected). Figure 8 of Ilbert et al. (2009) shows
  that the photo-z $1\sigma$ error derived from the PDFz is well
  representative of the photo-z accuracy. The median $1\sigma$ error
  is 0.02 for the full catalogue at $F_{3.6\mu m}>1\mu Jy$ and 0.08 in
  the redshift range $1.25<z<2$. We also showed in Figure 12 of Ilbert
  et al. (2009) that the photo-z accuracy is degraded at
  $i^+_{AB}>25.5$. Therefore, we take special care in limiting the
  contribution of these faint sources in our analysis (see
  \S\ref{opticalSel}).

\begin{figure}[htb!]
\includegraphics[width=7.9cm]{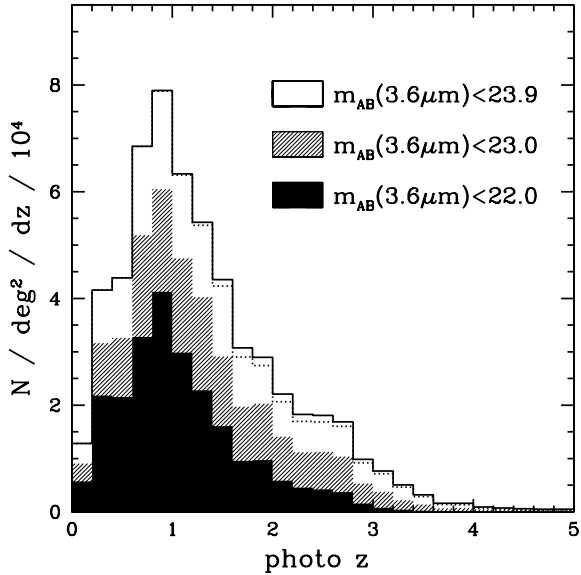}
\caption{Photometric redshift distributions for the 3.6$\mu m$
  selected sample ($m_{AB}(3.6\mu m)=-2.5 log(F)+23.9$ where $F$ is
  the flux in $\mu Jy$). The dotted line corresponds to the
    redshift distribution without the galaxies not detected in
    optical.\label{distz}}
\end{figure}

\subsection{Identification of the 3.6$\mu m$ source counterparts}\label{match}

We cross-matched the 3.6$\mu m$ and photo-z catalogues by taking the
closest counterpart within a radius of 1$^{\prime\prime}$. The
distances of the first and second closest optical counterpart have a
median value of 0.2$^{\prime\prime}$ and 2.5$^{\prime\prime}$,
respectively. The two distributions intersect at
1$^{\prime\prime}$. Therefore, we adopted a matching distance of
1$^{\prime\prime}$ which is a good compromise to detect the maximum of
optical counterparts and limits the risk of wrong
identification. Still, 2.6\% of the IRAC sources have two possible
optical counterparts in less than 1$^{\prime\prime}$.  In order to
estimate the probability of having identified the wrong optical
counterpart, we multiplied the probability of having the right
counterpart by the probability of having another optical source at a
lower distance (using the distance distributions of the first and
second closest counterpart, respectively). We obtained that the
probability of having identified the wrong optical counterpart is
0.1\%. We found a similar probability of 0.4\% using simulations.

We identified 8507 3.6$\mu m$ sources without optical counterpart
(about 4\% of the IRAC catalogue). Most of these sources are extremely
faint at 3.6$\mu m$, without counterpart in the K-band selected
catalogue (McCracken et al. 2009). This sample includes also a
significant fraction of fake detections created by the residual of the
muxbleed correction (Surace et al. 2005). Still, we were able to
identify 2714 IRAC sources which are clearly non-detected in optical
and are detected in the K-band selected catalogue. These sources can
be $z>1.5$ quiescent systems. Therefore, we included them in our
analysis. We measured a photo-z for these sources using NIR and IRAC
data. An upper-limit was set in $i^+$ since this band was used for
galaxy detection in Capak et al. (2009). The averaged redshift of this
population is $z \sim 2.9$. 93\% and 77\% of these sources are at
$z>1.5$ and $z>2$, respectively. In any case, the impact of this
galaxy population on our analysis is low since the stellar mass limits
are set to ensure a low fraction of sources with $i^+_{AB}>25.5$ in
the stellar mass sample (see \S\ref{opticalSel}).

Finally, we removed all of the sources flagged as star or AGN. Stars
were removed from the sample by comparing the $\chi^2$ evaluated for
both the galaxy templates and stellar templates (see \S 3.6 of Ilbert
et al. 2009). The 1,887 sources (1\% of the total sample) detected
with XMM-COSMOS (Hasinger et al. 2007, Brusa et al. 2007, Salvato et
al. 2009) were removed from the sample since their optical emissivity
is likely dominated by an AGN. \\

To summarize, this study is based on the S-COSMOS 3.6$\mu m$ selected
catalogue which is 50\% complete at 1 $\mu Jy$. We cross-match this
catalogue with the full optical and photo-z catalogue using a match
distance of 1$^{\prime\prime}$. The photo-z accuracy is as good as
$\sigma_{(z_{phot}-z_{spec})/(1+z_{spec})}=0.008$ at $i^+_{AB} <22.5$.
The final sample (after having removed the stars, the XMM sources, the
masked areas and the objects without optical counterparts) contains
196,000 galaxies at $F_{3.6\mu m} > 1 \mu Jy$ over an effective area
of 1.73-deg$^2$. Figure \ref{distz} shows the redshift distributions
for the $F_{3.6\mu m} > 1 \mu Jy$ selected sample with a median
redshift of $z \sim 1.1$.

\section{Morphological and spectral galaxy classifications}\label{classification}

In this paper, we study the galaxy stellar mass functions per
morphological and spectral type. This section presents the criteria
used to define the various galaxy types: E/S0 and Spi/Irr galaxies
based on their morphology ; three spectral classes (``quiescent'',
``intermediate activity'' and ``high activity'' galaxies) using
best-fit templates.

\subsection{Morphological classification}\label{morpho}

We used the high resolution HST/ACS images (Koekemoer et al. 2007) to
perform a morphological classification of our galaxy sample. The
images in the F814W filter reach a depth of 27.8 mag for a point
source at $5\sigma$. We adopted two independent morphological
classifications to separate E/S0 and Spi/Irr galaxies.

The first classification is based on the Gini (G) and concentration
(C) parameters measured by Abraham et al. (2007) (hereafter G-C
classification). The Gini parameter measures the inequality with which
the light of a galaxy is distributed among its constituent
pixels. Like Capak et al. (2007), the galaxies with $G>0.43$ were
considered E/S0 galaxies. In addition, we rejected from the E/S0
sample the galaxies with a concentration parameter smaller than 0.3
(Ilbert et al. 2006a).

\begin{figure}[htb!]
\includegraphics[width=7.9cm]{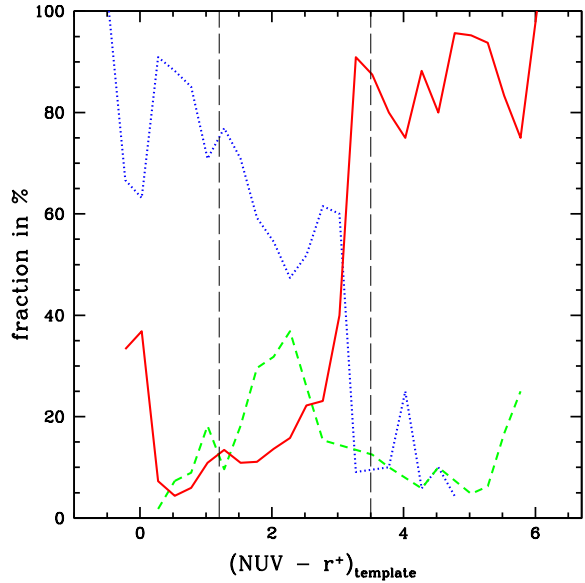}
\caption{Distribution of the unextincted rest-frame color $( {\rm NUV}
  -r^+ )_{template}$ of 1,500 visually classified galaxies (late
  spiral and irregular: blue dotted line; early spiral: green dashed
  line; E/S0: red solid line). The vertical dashed lines show the
  separation between the ``quiescent'', ``intermediate activity'' and
  ``high activity'' galaxies.\label{visual}}
\end{figure}

\begin{figure*}[htb!]
\includegraphics[width=17.5cm]{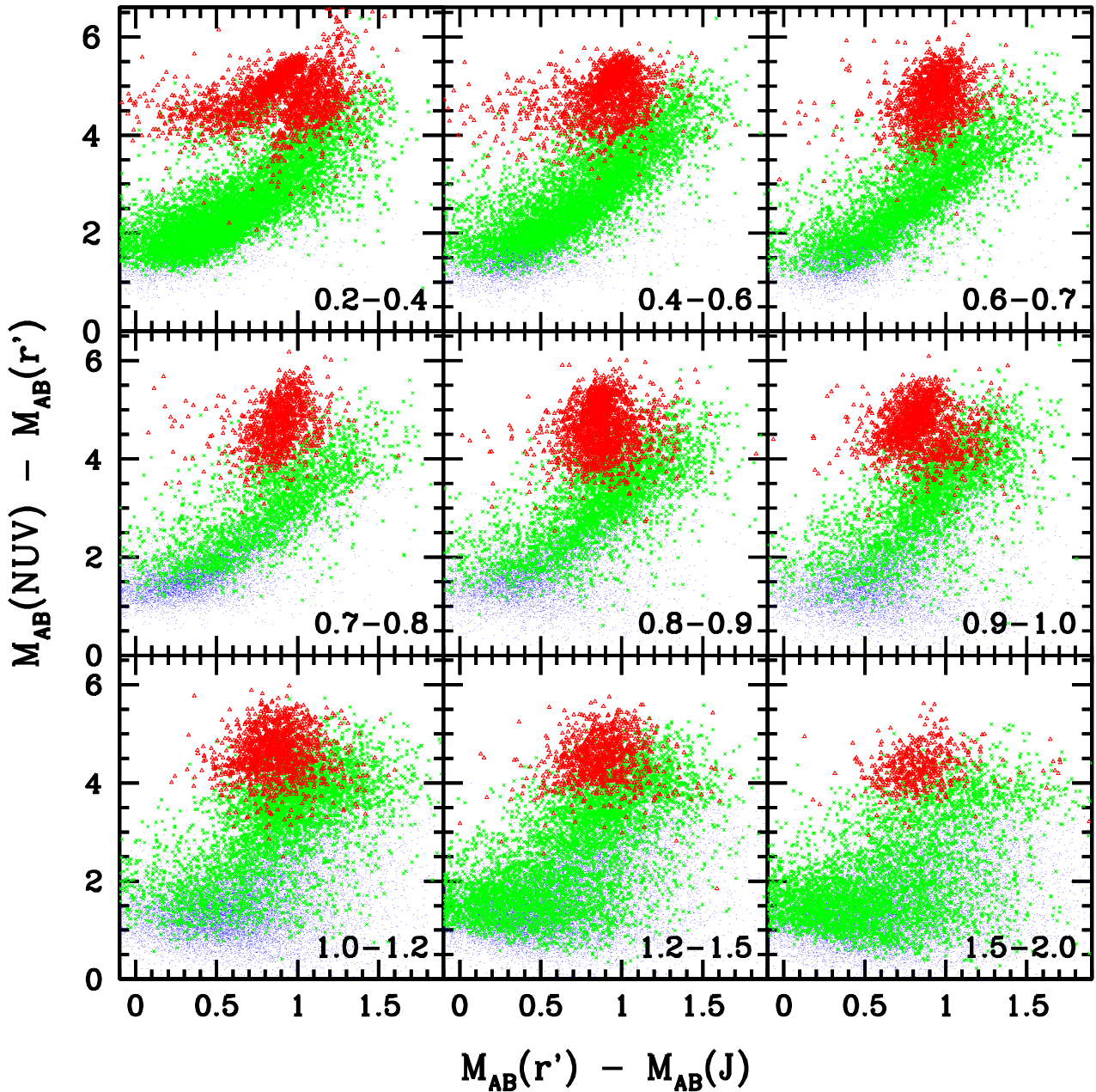}
\caption{Rest-frame colors $M({\rm NUV}) - M(r^+) $ versus $M(r^+) -
  M(J)$ (not corrected for dust reddening) from $z=0.2$ (top left
  panel) to $z=2$ (bottom right panel). The red open triangles, green
  crosses, and blue points are the galaxies selected as ``quiescent'',
  ``intermediate activity'' and ``high activity'', respectively, on
  the basis of their unextincted rest-frame color $( {\rm NUV} -r^+
  )_{template}$. \label{bimoColor}}
\end{figure*}

The second classification was performed by Cassata et al. (2009, in
prep.) (hereafter C09 classification). The structural parameters are
measured using a ``quasi-Petrosian'' image tresholding technique
(Abraham et al. 2007). This classification includes Gini,
Concentration, Asymmetry, and M20 (e.g. Lotz et al. 2004). The
multi-dimensional parameter space is automatically converted into an
E/S0 and Spi/Irr classification by matching these parameters with those
of a training sample of 250 visually classified galaxies (50 galaxies
per 0.5 mag bin out to $i^+_{AB}<24$).

The E/S0 selection performed by C09 is more conservative than the G-C
classification (i.e. less contaminated by spiral galaxies), but is
likely to be more incomplete. Indeed, less than 1\% of the E/S0
sources from C09 are not identified as E/S0 with the G-C parameters,
while 33\% of the E/S0 sources identified with the G-C parameters are
not identified by C09 (at $log({\cal M})>10$ and $z<1.2$).

\begin{figure}[htb!]
\includegraphics[width=7.9cm]{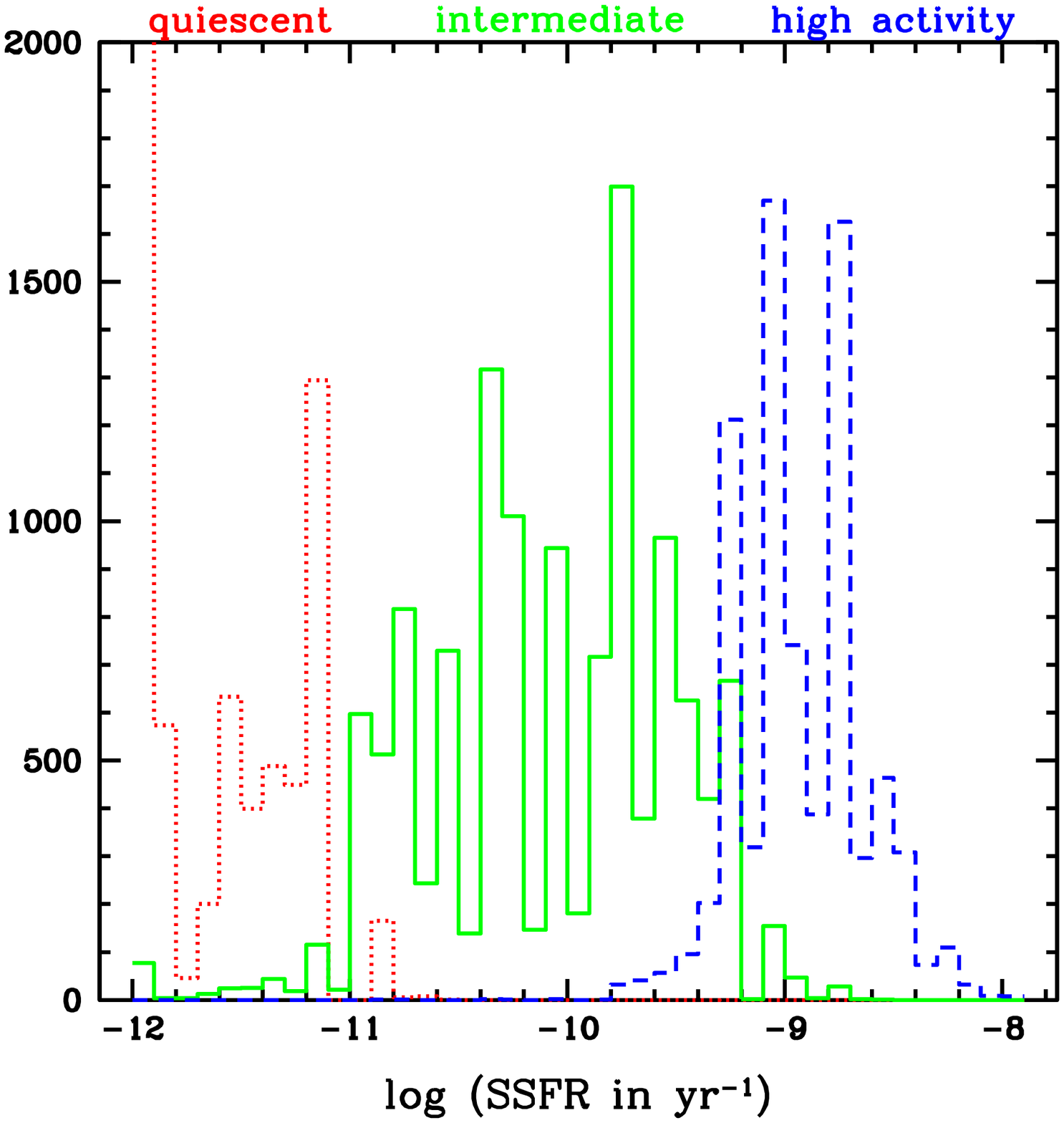}
\caption{Distribution of the Specific SFR (SSFR) for the ``quiescent'' (red
  dotted line), ``intermediate activity'' (green solid line) and
  ``high activity'' (blue dashed line) galaxies (at $0.2<z<1.2$ and
  $log({\cal M})>10$) (see \S\ref{template}). \label{SSFR}}
\end{figure}

\subsection{Spectral classification}\label{template}

A set of SED templates was generated using the Bruzual \& Charlot
(2003) (BC03) package and fitted to the multi-color data (see
section.\ref{stellarmass}). The extinction is added as a free
parameter in the fit. We used the unextincted rest-frame colors $({\rm
NUV} -r^+)_{template}$ of the templates to define three spectral
classes: {\it a)} the ``quiescent'' galaxies with $( {\rm NUV} -r^+
)_{template}>3.5$; {\it b)} the ``intermediate activity'' galaxies
with $1.2<( {\rm NUV} -r^+ )_{template}<3.5$; {\it c)} the ``high
activity'' galaxies with $( {\rm NUV} -r^+ )_{template}<1.2$.

Figure \ref{visual} shows the $( {\rm NUV} -r^+ )_{template}$
distribution of a sample of 1,500 galaxies that we visually classified
as E/S0, early spiral, late spiral or irregular. The 1,500 galaxies
were selected to provide an unambiguous visual classification
(isolated and bright galaxies) but were not selected to be
  statistically representative of the $3.6\mu $m sample. A cut at $(
{\rm NUV} -r^+ )_{template}>3.5$ isolates well the E/S0 galaxies. The
``intermediate activity'' class ($1.2<( {\rm NUV} -r^+
)_{template}<3.5$) includes most of the visually selected early spiral
galaxies but is strongly contaminated by late spiral and irregular
galaxies.

Figure \ref{bimoColor} shows a slightly modified version of the
color-color selection technique ($M_U-M_V$ versus $M_V-M_J$) proposed
by William et al. (2008). We used the color ${\rm NUV} -r^+$ instead
of $U-V$ since this color is a better indicator of the current versus
past star formation activity (e.g. Martin et al. 2007, Arnouts et
al. 2007).  The rest-frame colors were computed as described in
appendix \ref{absmag} and are not corrected for internal dust
attenuation (by contrast with $({\rm NUV} -r^+ )_{template}$ which is
corrected for dust attenuation). A red clump appears clearly from
$z=0.2$ out to $z=2$. This clump is mostly composed of ``quiescent''
galaxies. Therefore, our ``quiescent'' population is similar to the
red clump population selected by William et al. (2008). The galaxies
with a red ${\rm NUV} -r^+>4.5$ rest frame color are well separated
into a ``quiescent'' population with ${\rm r^+} - J<1.2$ (red clump)
and a dust-extincted star-forming population with ${\rm r^+} - J>1.2$.

Finally, we show in Figure \ref{SSFR} that each spectral class
corresponds to a range of Specific SFR (SSFR), computed as the
instantaneous SFR from the best-fit template divided by the stellar
mass.\\

To summarize, we used the HST/ACS images to separate E/S0 and Spi/Irr
galaxies using two morphological classification methods (C09 and
G-C). We also defined three spectral classes on the basis of the
best-fit templates which are ``quiescent'', ``intermediate activity''
and ``high activity'' galaxies. These three spectral classes are well
separated in ranges of SSFR. We showed that our ``quiescent'' population
matches well with the red clump galaxies found by Williams et
al. (2008) and is consistent with an E/S0 population selected
morphologically.

\begin{figure*}[htb!]
\begin{tabular}{c c}
\includegraphics[width=7.9cm]{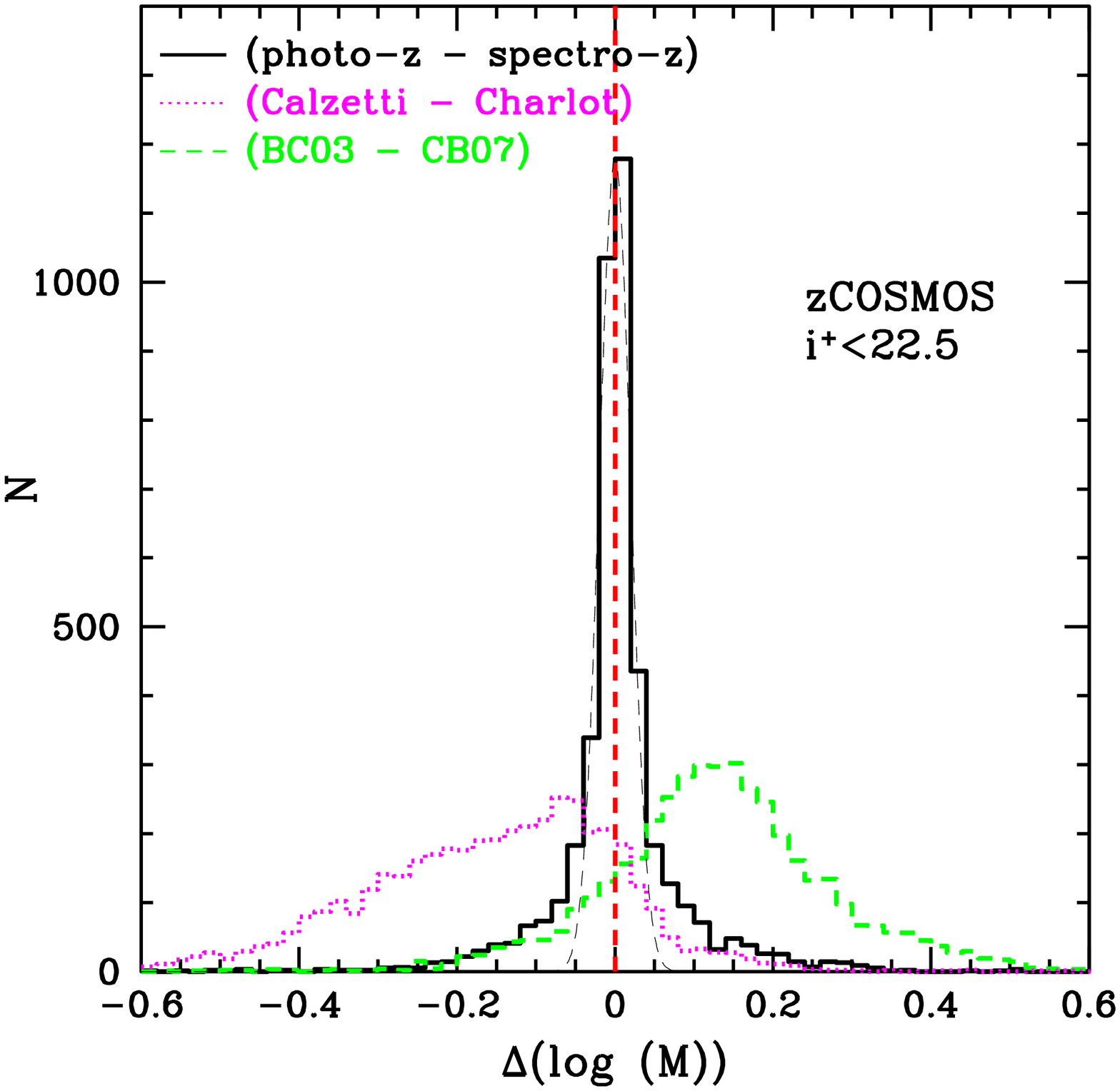} &
\includegraphics[width=7.9cm]{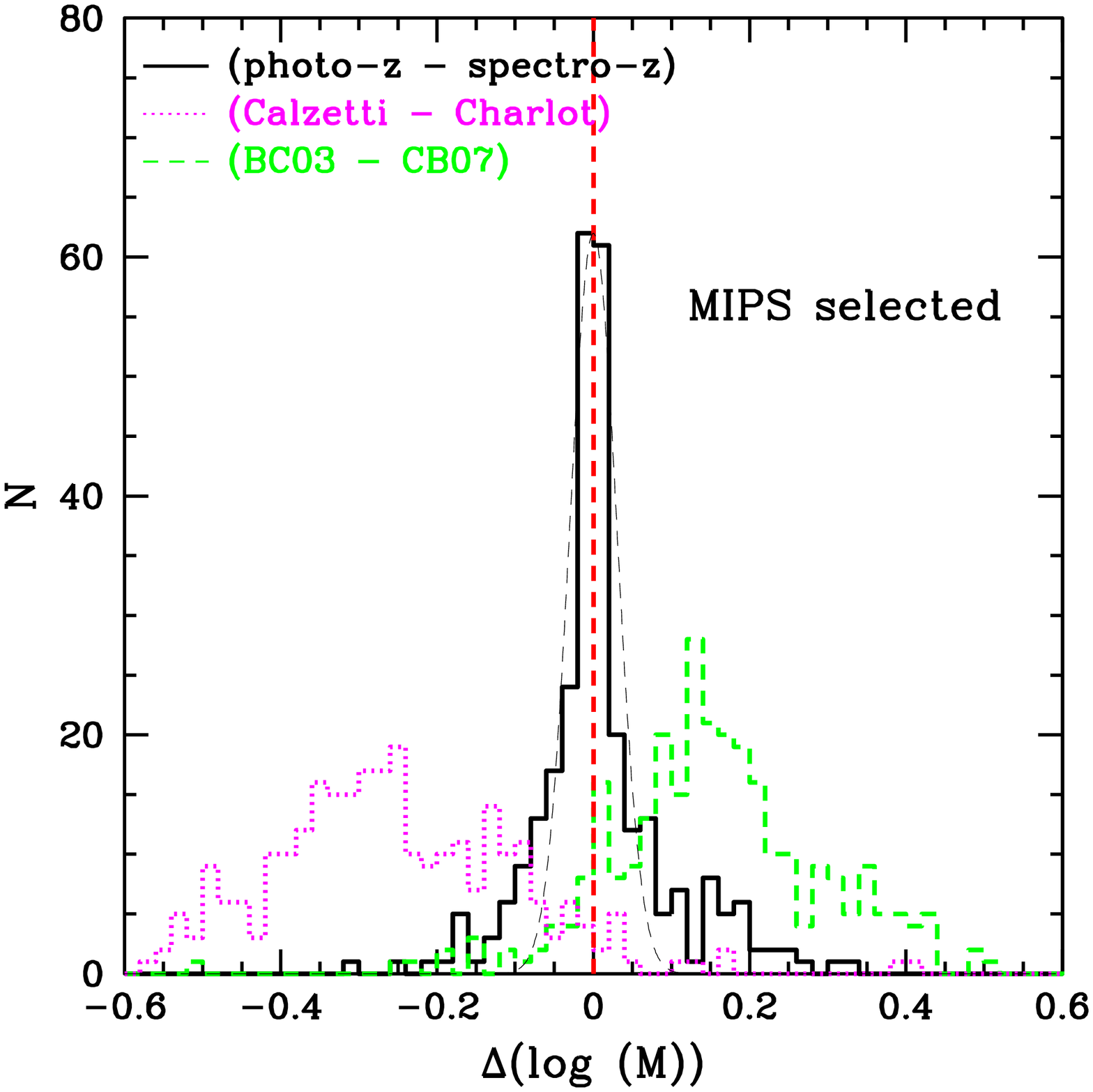}
\end{tabular}
\caption{The black solid line histograms show the difference between the stellar masses computed with photo-z and spectro-z. We used the zCOSMOS spectroscopic sample selected at $i^+_{AB}<22.5$ (Lilly et al. 2009, in prep.) in the left panel and the spectro-z of infrared selected sources from Kartaltepe et al. (2009, in prep.) in the right panel. The thin black dashed lines are gaussian distributions with $\sigma=0.02$ (left panel) and $\sigma=0.03$ (right panel). The green dashed lines show the difference between the stellar masses computed with BC03 and Charlot \& Bruzual (2007, private communication). The magenta dotted lines show the differences between the stellar masses computed using the Calzetti et al. (2000) and Charlot \& Fall (2001) extinction laws. The redshifts were set to the spectro-z values in the two last cases. Systematic uncertainties due to the models dominate the errors introduced by the photo-z, at least in the magnitude/redshift range explored with our spectroscopic samples. \label{hist_SM_zp_zs}}
\end{figure*}

\begin{table}[htb!]
\begin{center}
\begin{tabular}{c | c | c} 
\hline & & \\
$\tau$ (Gyr) & E(B-V) & $Z$ \\
\hline & &  \\
 0.1         &      0      & 0.02 ($Z_{\Sun}$) \\
 0.3         &     0.1     & 0.008  \\
   1         &     0.2     &        \\
   2         &     0.3     &        \\
   3         &     0.4     &        \\
   5         &     0.5     &        \\
  10         &             &        \\
  15         &             &        \\
  30         &             &        \\
\hline  
\end{tabular}
\caption{Parameters used to generate the SED templates with the BC03 package.}
\label{tab0}
\end{center}
\end{table}

\section{The galaxy stellar mass sample}\label{method}

In this section, we describe the method used to measure galaxy stellar
masses and the galaxy stellar mass function.

\subsection{Technique used for estimating stellar masses}\label{stellarmass}

We used stellar population synthesis (SPS) models to convert
luminosity into stellar mass (e.g. Bell et al. 2003, Fontana et
al. 2004). The stellar mass is the factor needed to rescale the
best-fit template (normalized at one solar mass) for the intrinsic
luminosities. The SED templates were generated with the stellar
population synthesis package developed by (BC03). We assumed an
universal IMF from Chabrier (2003) and an exponentially declining star
formation history $SFR \propto e^{-t/\tau}$ ($\tau$ in the range 0.1
Gyr to 30 Gyr). The SEDs were generated for a grid of 51 ages (in the
range 0.1 Gyr to 14.5 Gyr). Dust extinction was applied to the
templates using the Calzetti et al. (2000) law ($E(B-V)$ in the range
0 to 0.5). We used models with two different metallicities. The
parameters used to generate the SED templates are listed in Table
\ref{tab0}. Following Fontana et al. (2006) and Pozzetti et
al. (2007), we imposed the prior $E(B-V)<0.15$ if $age/\tau >4$ (a
significant extinction is only allowed for galaxies with a high SFR).

We introduced the fluxes measured at 24 $\mu m$ with the Spitzer
/MIPS instrument (Aussel et al. 2009, in prep.) as a constraint in the
template fitting procedure, as detailed in appendix \ref{priorSFR};
however, the 24 $\mu m$ constraint had little effect on the derived
stellar masses. The dispersion between the stellar masses with and
without using this constraint is 0.014 dex, and only 1\% of the sample
differs by more than 0.2 dex.

All the available bands were used to compute the stellar masses (broad
bands as well as medium bands). The IRAC data were included in the
fit. We compared the stellar masses computed with and without IRAC. We
obtained a dispersion of 0.02, 0.02, 0.04, 0.11 dex at z=0.2-0.5,
z=0.5-1, z=1-1.5, z=1.5-2, respectively. At $z<1.5$, the IRAC data
have a small impact on the stellar masses, negligible in comparison to
the 0.2 dex uncertainty expected in the stellar mass estimate
(e.g. Pozzetti et al. 2007, Longhetti \& Saracco 2008). At $z>1.5$,
including the IRAC data modifies significantly the stellar masses. The
IRAC data are necessary since the $K$ band doesn't probe anymore the
NIR rest-frame wavelength range.

\subsection{Systematic uncertainties in the stellar mass estimate}\label{teststellarmass}

We quantified how the stellar mass accuracy is affected by the use of
photo-z rather than spectro-z. Figure \ref{hist_SM_zp_zs} shows the
difference between the stellar masses computed with the photometric
and spectroscopic redshifts. We used two spectroscopic samples: the
zCOSMOS bright spectroscopic sample selected at $i^+_{AB}<22.5$ (Lilly
et al. 2009, in prep.) and a spectroscopic follow-up of $24 \mu m$
selected sources (median flux $F_{24\mu m} \sim 140\mu Jy$) by
Kartaltepe et al. (2009, in prep.). The infrared follow-up supplements
very well the zCOSMOS spectro-z since the former sample is fainter
($18< i^+_{AB}< 25$ with 43\% of the sources being fainter than
  $i^+_{AB}=22.5$) and extends out to $z\sim 1.5$ (median redshift of
$z\sim 0.74$). We found a median difference smaller than 0.002 dex
between the photo-z and the spectro-z stellar masses for both
samples. Therefore, no systematic offsets appear to be introduced by
the use of our photo-z. The dispersion between the two estimates is
smaller than $\sim 0.03$ dex. This dispersion is 10$\times$ smaller
than the systematic uncertainties expected in the stellar mass
estimate (e.g. Pozzetti et al. 2007, Longhetti \& Saracco 2008), at
least in the magnitude/redshift space covered by the spectroscopic
samples.

The choice of extinction law impacts the fit of the template and
therefore the mass-to-light ratio. We computed the stellar masses
using both the Calzetti et al. (2000) and the Charlot \& Fall (2000)
extinction laws (the latter is included in BC03). The redshifts are
set to the spectro-z values. The median difference between the two
stellar mass estimates (Calzetti - Charlot \& Fall) is -0.14 dex with
a dispersion of 0.10 dex for the zCOSMOS sample (left panel of Figure
\ref{hist_SM_zp_zs}). This median difference reaches -0.27 dex
for the MIPS spectroscopic sample, showing that the systematic offset
is larger for massive galaxies with a high SFR. We
adopted the Calzetti et al. (2000) extinction law. This choice is
favored by a comparison between the SFR derived from the best-fit
template and the SFR measured from mid-infrared 24 $\mu m$ data, as
described in appendix \ref{priorSFR}.

The stellar mass estimate depends also on the assumed stellar
population synthesis model. We computed the stellar masses with an
upgraded version of the BC03 model including a better treatment of the
thermally pulsing asymptotic giant branch (TP-AGB) phase (Bruzual G.,
2007, Charlot \& Bruzual, 2007, private communication). Figure
\ref{hist_SM_zp_zs} shows a comparison of the stellar mass estimates
using the two versions of the models. We find a median difference of
0.13-0.15 dex and a dispersion of 0.09 dex. Pozzetti et al. (2007)
measured a difference of 0.14 dex between the stellar masses computed
with the BC03 and the Maraston (2005) models (which also includes
treatment of the TP-AGB stars), in agreement with our results. We used
the public version of BC03 for consistency with results from the
literature. However, the MF computed in this paper have also been
computed with the upgraded version of BC03, and our conclusions
remained unchanged.

\subsection{Method to estimate the mass function}\label{method}

We measured the stellar mass functions using the tool ALF ({\it
  Algorithm for Luminosity Function}, Ilbert et al. 2005). This tool
includes the STY parametric estimator (Sandage, Tammann \& Yahil 1979)
and three non-parametric estimators: the $1/V_{max}$ (Schmidt 1968),
$C^+$ (Lynden-Bell 1971 and Zucca et 1997) and the Step-Wise Maximum
Likelihood (SWML, Efstathiou 1988). A brief introduction of these
standard estimators is given in appendix \ref{ALF}.

We weighted each galaxy according to the completness of the
  3.6$\mu m$ catalogue (see section \ref{IRACcatalogue} and Figure
  \ref{completeness}). We attributed a weight to each source depending
  on the 3.6$\mu m$ flux. This weight is the inverse fraction of
  3.6$\mu m$ sources detected at this flux (e.g. a weight of 2 is
  given to the sources at $F_{3.6\mu m}\sim 1 \mu Jy$).

We also performed extensive simulations in order to propagate the
photo-z uncertainties into the mass function. A redshift probability
distribution function (PDFz) was attributed to each galaxy when we
measured the photo-z (see Ilbert et al. 2009). We created 20
catalogues by randomly picking a redshift within the PDFz of each
object.  The MFs were measured for each of the 20 catalogues in every
redshift bin and for each galaxy population. Finally, the dispersion
of the Schechter parameters was measured over the 20 realizations. We
added in quadrature the Poissonian errors and these errors induced by
the use of photo-z.

\begin{figure}[htb!]
\includegraphics[width=7.9cm]{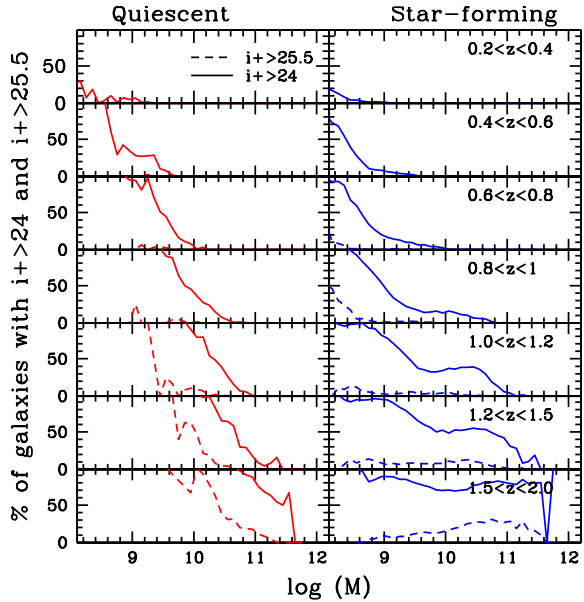}
\caption{Fraction of galaxies with an apparent magnitude fainter than $i^+=24$ (solid lines) and $i^+=25.5$ (dashed lines) as a function of stellar mass. This fraction is measured per redshift bin (top to bottom panels). The left and right panels correspond to the quiescent and star-forming galaxies, respectively. From this fraction, we defined the lowest stellar mass limit which ensures a maximum of 30\% of galaxies fainter than $i^+>24$ and $i^+>25.5$ in the lowest stellar mass bin of the MF. \label{MSsel_h}}
\end{figure}

\subsection{Considered stellar mass range for the MF estimate}\label{opticalSel}

The low mass limits considered for the MF estimates are set in order
to insure a complete and unbiased stellar mass sample with accurate
photo-z.

We defined the low stellar mass limits in order to reduce the fraction
of optically faint sources with low quality photo-z in the stellar
mass sample. Figure \ref{MSsel_h} shows the fraction of galaxies with
$i^+>25.5$ (dashed lines) as a function of the stellar mass. We
defined the stellar mass ranges in order to keep the fraction of
galaxies fainter than $i^+>25.5$ below an arbitrary limit of 30\%. We
set the limit to $i^+_{AB}=25.5$ since the photo-z are degraded at
fainter magnitudes (Ilbert et al. 2009). With this approach, the
lowest stellar mass bin of the MF has 30\% of its objects with lower
accuracy photo-z, and this fraction decreases rapidly in the higher
stellar mass bins. According to Abraham et al. (2007) and Capak et
al. (2007), the morphological classification is robust out to
$i^+_{AB}\sim 24$. When a morphological selection is applied, we
therefore adopted a limit at $i^+_{AB}<24$ (solid line in Figure
\ref{MSsel_h}) rather than $i^+_{AB}<25.5$.

Moreover, Ilbert et al. (2004) and Fontana et al. (2004) showed that
the MF estimators can be biased at low masses because galaxies with
different SEDs (and mass-to-light ratios) are not visible up to the
same stellar mass limit. This affects each MF estimator differently
(see Figure 4 in Ilbert et al. 2004). We restricted our MF estimate to
the stellar mass range where the 3 non-parametric estimators agree, to
limit the impact of such bias on our results.

The lowest mass limits considered for the MF estimate are given in
Table \ref{schechterRed} and Table \ref{schechterBlue} for each
redshift bin. In all cases, the lowest limit in mass has to be the
largest value imposed by the mix of galaxy type and the photo-z
limitation at $i^+_{AB}<25.5$.

\begin{figure}[htb!]
\includegraphics[width=7.9cm]{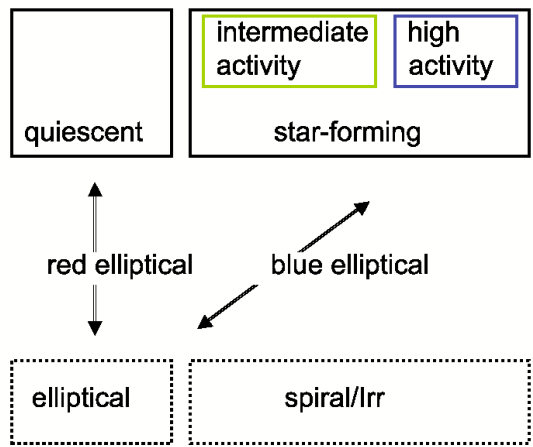}
\caption{Schematic view of the different classifications used in this paper. The top and bottom panels are the spectral and morphological classifications, respectively. \label{Figclassification}}
\end{figure}

\begin{figure*}[htb!]
\includegraphics[width=17.cm,height=17.5cm]{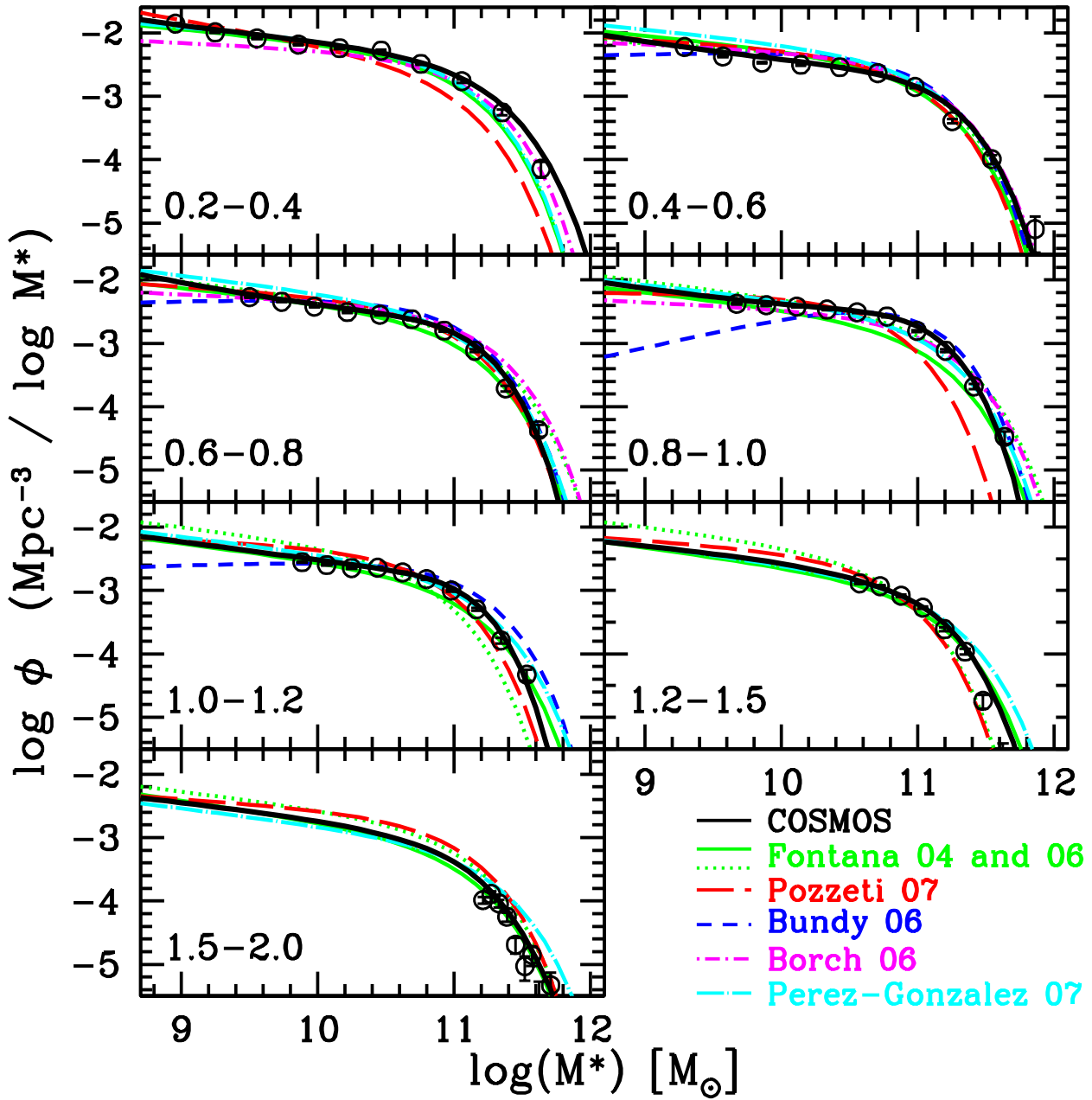}
\caption{Total MF (no separation by type). The open circles are the non-parametric estimates of the MF ($1/V_{max}$). The solid black lines are the sum of the ``quiescent'', ``intermediate activity'', and ``high activity'' MFs as taken from Table \ref{schechterRed} and Table \ref{schechterBlue}. The total MFs are compared with data from the literature (all MFs are converted into a Chabrier IMF). \label{MFglob}}
\end{figure*}

\section{Mass Function of the total sample}\label{MFtot}

We first analyze the total MF (no cut by morphological or spectral
type) and compare it with data from the litterature. In the next
sections, we will study the MF per morphological and spectral type
(Figure \ref{Figclassification} summarizes the different
classifications used hereafter.)\\

Figure \ref{MFglob} shows the estimate of the total MF. The
non-parametric MF estimate (open black circles) shows a small
turn-over at $log({\cal M})<10$ and $z<0.8$ which can not be
reproduced well with a Schechter function (Schechter 1976, see
appendix \ref{ALF}). This turn-over is also detected in Pozzetti et
al. (2009) and Drory et al. (2009). Therefore, we do not give a
Schechter parametrization of the total MF. A parametrization can be
retrieved by summing the Schechter fits of the ``quiescent'',
``intermediate activity'', and ``high-activity'' galaxies given in
Table \ref{schechterRed} and Table \ref{schechterBlue}. This sum
(solid black curves in Figure \ref{MFglob}) provides a parametrization
in excellent agreement with the non-parametric estimate.

We compared the total MF and data from the literature (all MFs are
converted into a Chabrier IMF and to the same cosmology). In general,
we find excellent agreement between the different MFs out to
$z=2$. The offsets between the high-mass exponential cutoffs (i.e. the
sharp decline of the density above the characteristic stellar mass
${\cal M}^*$) are smaller than 0.2 dex. Combined differences due to
cosmic variance and methodology used to measure the stellar
masses\footnote{For instance, Bundy et al. 2006 used the Charlot \&
Fall 2001 extinction law, Pozzetti et al. 2007 did not allow sub-solar
metallicities, and Borch et al. 2006 used the PEGASE stellar
population synthesis package from Fioc \& Rocca-Volmerange 1997 and no
NIR data.} are consistent with differences of 0.2 dex (see
\S\ref{teststellarmass}).

\begin{figure*}[htb!]
\includegraphics[width=17.cm,height=17cm]{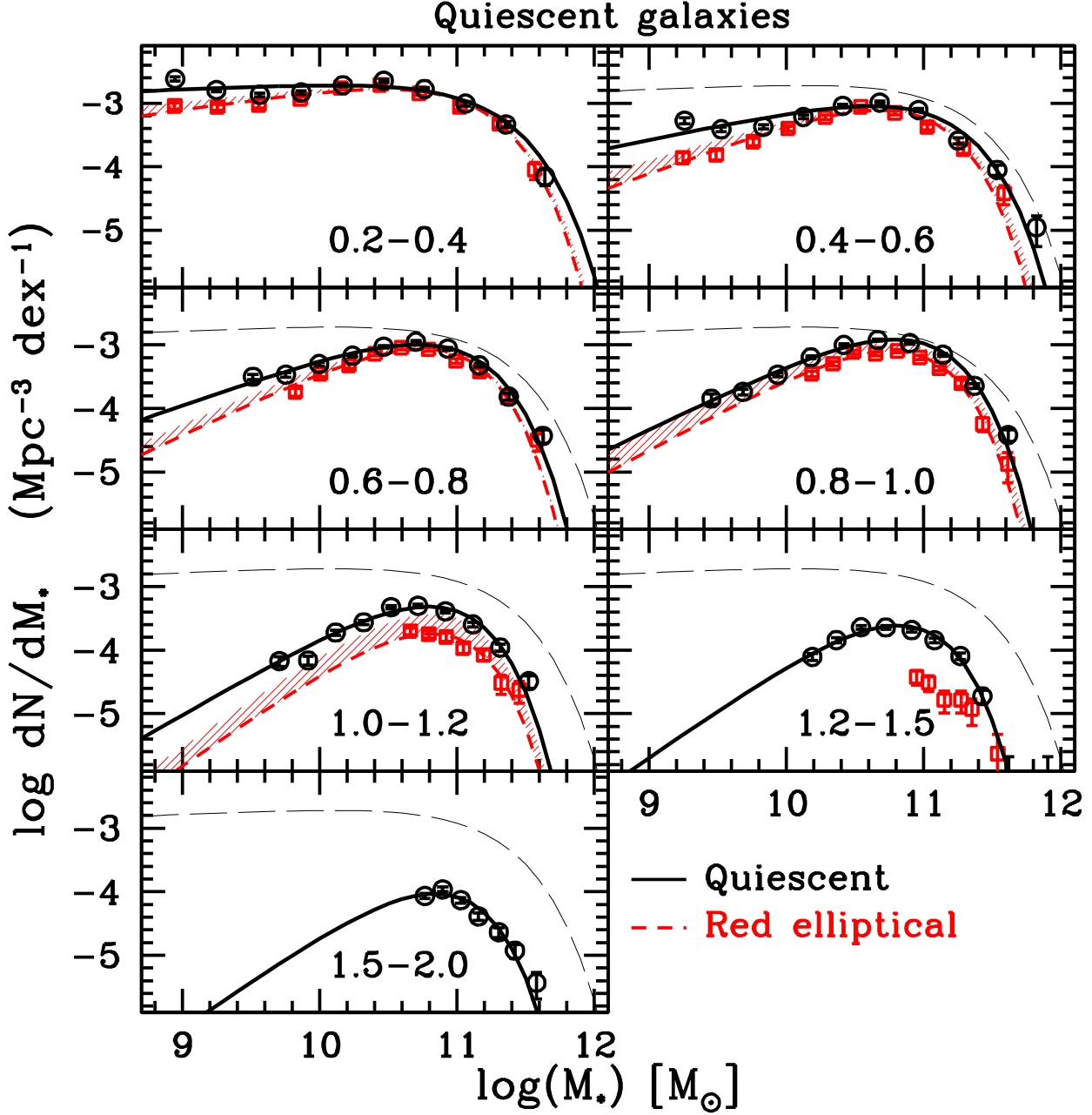}
\caption{Evolution of the MFs for ``quiescent''
  galaxies from $z=0.2-0.4$ (top left panel) to $z=1.5-2$ (bottom left
  panel). The ``quiescent'' galaxies are shown by black open circles
  and solid lines. The lower and upper envelopes of the shaded area
  are the red elliptical classified using C09 and G-C,
  respectively. The black long-dashed line is the quiescent MF
  estimated at $z=0.2-0.4$; it is shown in each panel to serve as a
  reference. \label{MF_TBR}}
\end{figure*}

\section{Stellar Mass assembly of quiescent and elliptical galaxies}\label{MFearly}

In this section, we present the MF for galaxies having a ``quiescent''
spectral type. From the tight correlation seen in the local Universe
between morphology and colors, the quiescent galaxies are expected to
preferentially have an elliptical morphology. However, the correlation
between color and morphology is not perfect (Bell et al. 2008) and we
need to quantify how this relation evolves with redshift. Therefore,
we also derived the MF of quiescent galaxies having an elliptical
morphology (``red elliptical'', see Figure \ref{Figclassification}).

\subsection{MF of quiescent galaxies }\label{MF_ell}

\begin{figure}[htb!]
\includegraphics[width=7.9cm]{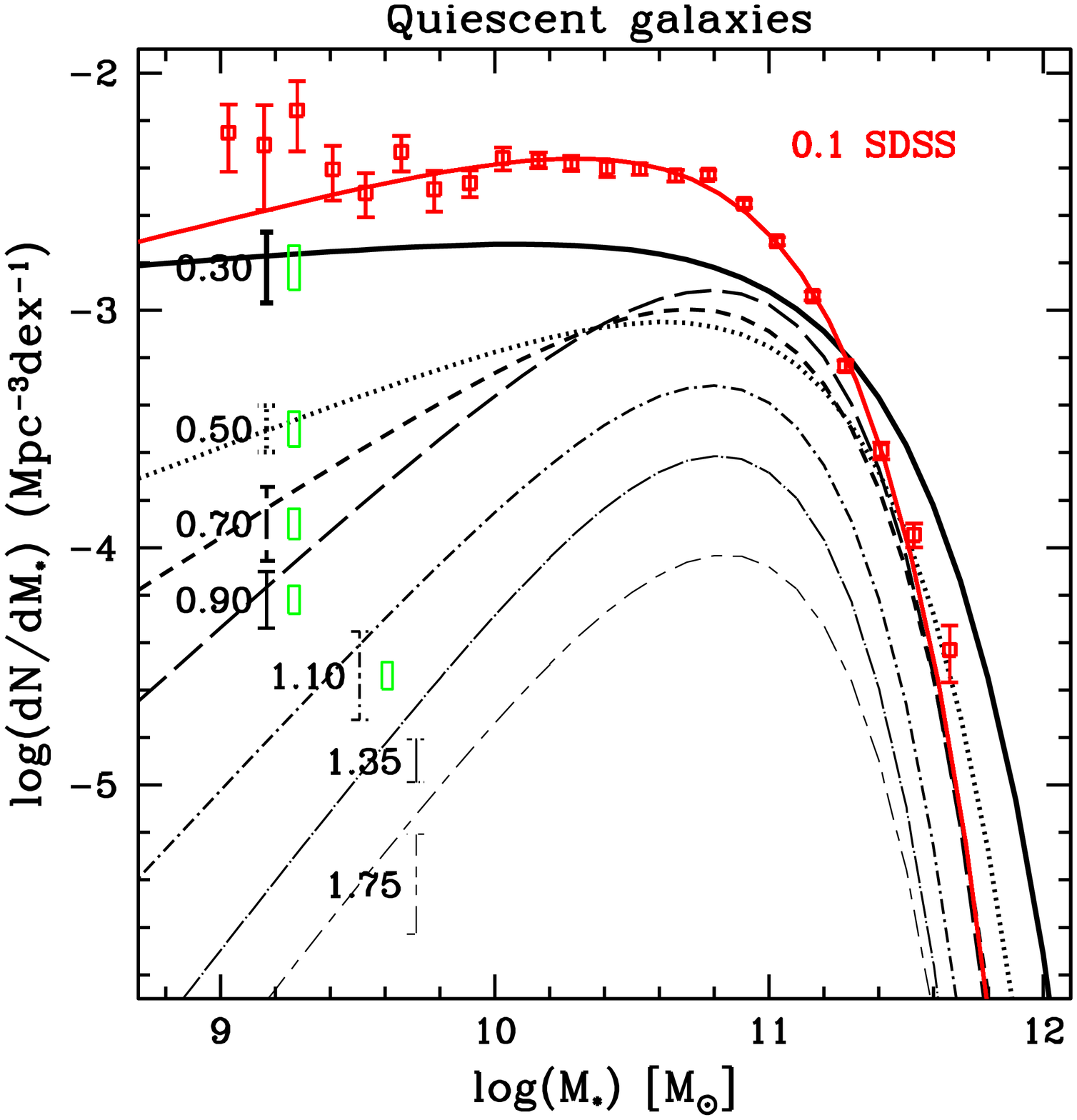}
\caption{MF of the ``quiescent'' galaxies from $z=1.5-2$ to
  $z=0.2-0.4$. The numbers on the MFs are the redshifts. The green
  vertical boxes are an estimate of the cosmic variance by Scoville et
  al. (2007) for halo mass ranges of $10^{13}-10^{14}{\cal
    M}_\Sun$. The black vertical lines correspond to the two extreme
  MFs in 4 sub-fields obtained by splitting the 2-deg$^2$ into four
  quadrants each of 0.5-deg$^2$. The red solid line and points are the
  local MF measurements for red galaxies by Bell et al. (2003) at
  $z\sim 0.1$.  \label{MF_quiescent}}
\end{figure}

Figure \ref{MF_TBR} shows the MF of the quiescent galaxies in the
range $z=0.2-2$. We find that: {\bf i)} the density of ``quiescent''
galaxies more massive than $log({\cal M})>11$ increases by a factor of
$\sim 14$ between $z=1.5-2$ and $z=0.8-1$; {\bf ii)} this evolution
slows down significantly after $z<1$ and the high-mass exponential
cutoff does not increase by more than 0.2 dex at $z<1$; {\bf iii)} the
density of quiescent galaxies increases at intermediate mass between
$z=0.8-1$ and $z=0.2-0.4$ (e.g. by a factor of 4.4 at $log({\cal
  M})\sim 10$). In Figure \ref{MF_quiescent}, we have over-plotted our
results and the local measurement performed by Bell et al. (2003). We
find consistent evolutionary trends when we compare our data to the
local measurement\footnote{The local MF is computed with the code
  PEGASE. The local MF could be shifted by +0.06 dex to match our
  stellar masses computed with the BC03 code (Rettura et al. 2006).}:
the local density is higher at intermediate masses ($9<log({\cal
  M})<11$) and the local exponential cutoff is consistent within 0.2
dex with the values obtained at $0.2<z<1$.

\begin{figure}[htb!]
\includegraphics[width=7.9cm]{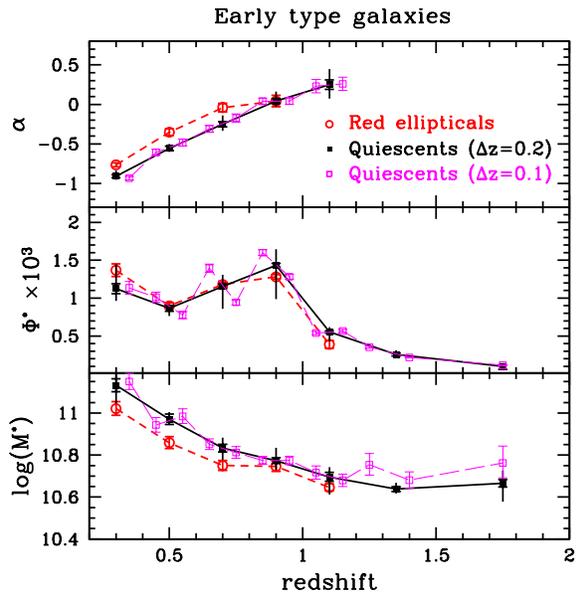}
\caption{Evolution of the Schechter parameters as a function of
  redshift (black solid lines: quiescent; red dashed lines: red
  ellipticals). The long dashed magenta lines show the Schechter
  parameters measured in smaller redshift bins of $\Delta z=0.1$ for
  the quiescent galaxies. From the top to the bottom: evolution with
  redshift of the slope, the normalization and the characteristic
  stellar mass, respectively. The vertical lines for the quiescent
  galaxies represent the extreme values in four sub-fields of
  0.5-deg$^2$ after having divided the COSMOS field into four equal
  quadrants. \label{all_ell}}
\end{figure}

We analyze the evolution of the best-fit Schechter parameters of the
quiescent MF (Table \ref{schechterRed}). The top panel of Figure
\ref{all_ell} shows a continuous steepening of the slope $\alpha$ with
time. This steepening reflects the rapid density increase of the
low/intermediate mass galaxies. In the middle panel, the normalization
$\Phi^*$ of the quiescent MF is shown to increase by a factor of 15
from $z=1.5-2$ to $z=0.8-1$. The rapid increase of $\Phi^*$ is no
longer detected from $z=0.8-1$ to $z=0.2-0.4$ where $\Phi^*$ remains
approximately constant. Some fluctuations (a factor of $\sim 2$)
appear in this redshift range when we reduce the size of the redshift
bins to $\Delta z=0.1$, which is consistent with cosmic variance
(Scoville et al. 2007). Finally, the characteristic stellar mass
${\cal M}^*$ increases by 0.3-0.4 dex between $z \sim 1$ and $z \sim
0.3$ (bottom panel).

\begin{figure}[htb!]
\includegraphics[width=7.9cm]{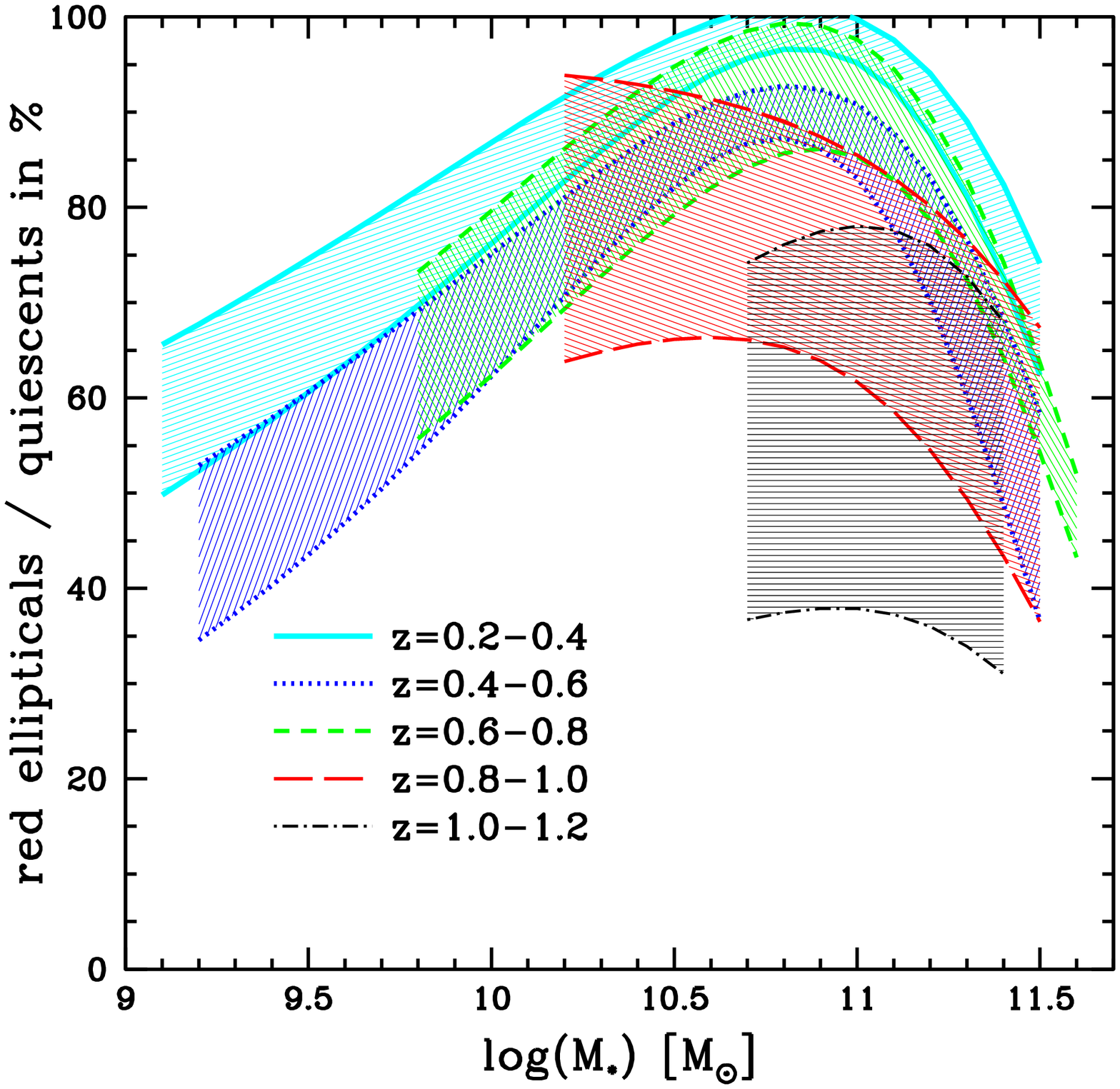}
\caption{Fraction (in \%) of ``quiescent'' galaxies with an elliptical morphology. The fraction is obtained by dividing the MF of the red ellipticals by the MF of the quiescent galaxies. The upper and lower limits are obtained using the G-C and C09 morphological classifications, respectively. \label{compositionMF}}
\end{figure}

\subsection{MF of red elliptical galaxies }\label{MF_ell}

Figure \ref{MF_TBR} shows the MF of the red ellipticals (quiescent
with an elliptical morphology) galaxies in the range $z=0.2-1.2$. We
find a similar evolution for the red ellipticals as for the quiescent
galaxies. Between $z=0.8-1$ and $z=0.2-0.4$, the density of red
ellipticals increases by a factor of 4-5.4 at intermediate mass
($log({\cal M})\sim 10$). By contrast, their density increases only by
1.7-2 at high mass ($log({\cal M})>11$). Therefore, the most massive
red elliptical galaxies show little evolution at $z<1$ while their
density still increases at low/intermediate masses.

The ratio between the red elliptical and quiescent MFs is plotted in
Figure \ref{compositionMF}. This ratio quantifies the fraction of
quiescent galaxies with an elliptical morphology (and the
complementary information about the fraction of Spi/Irr with quenched
star formation). At high mass, $log({\cal M})\sim 11$, the fraction of
quiescent galaxies with an elliptical morphology is greater than 90\%
(80\%) at $z<0.8$ for a G-C (C09) classification. This fraction has a
maximum at $10.5<log({\cal M})<11$ and decreases continously toward
low masses reaching 60\% (40\%) at $log({\cal M}) \sim 9.5$. This
  fraction seems also to decrease at really high mass ($log({\cal
    M})\sim 11.5$). However, we caution the reader that the constraint
  on the MF at such high masses relies on few galaxies. 

The fraction of quiescent galaxies with an elliptical morphology
decreases at $z>0.8$ (Figure \ref{compositionMF}). This decrease is
seen with both classifications. The decrease is much faster if we
consider the C09 classification which is more conservative in
selecting pure elliptical galaxies. This trend could show that: the
fraction of Spi/Irr with a quenched SFR increases with redshift ; the
``red and dead'' local elliptical galaxies didn't have fully acquired
their elliptical morphology at $z>0.8$. This result is discussed in
Bundy et al. (2009) and Oesch et al. (2009).

\begin{figure}[htb!]
\includegraphics[width=7.9cm]{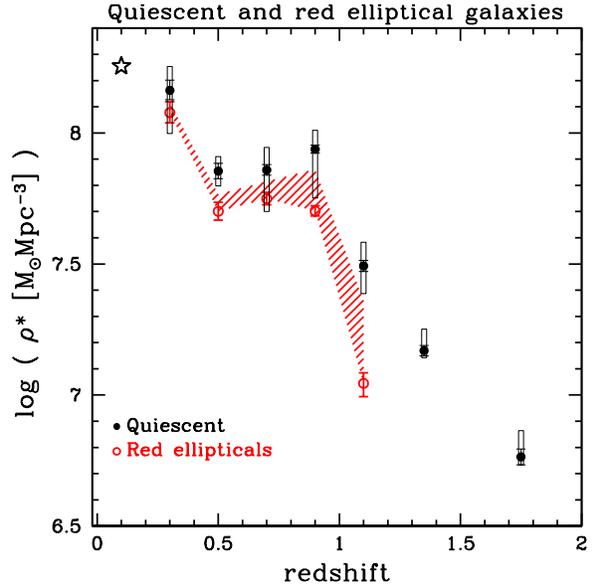}
\caption{Stellar mass density of quiescent galaxies (black solid
  circles) and red ellipticals (red open circles). The lower and upper
  envelopes of the shaded area correspond to the stellar mass density
  computed using the C09 and G-C classifications, respectively. The
  open star is the local measurement by Bell et al. (2003). The
  vertical boxes for the quiescent correspond to the two extreme
  values measured in 4 quadrants of 0.5-deg$^2$.
\label{LD_Ell}}
\end{figure}

\subsection{Stellar mass density of quiescent and red elliptical galaxies}\label{LDearly}

The stellar mass density quantifies the total stellar mass locked up in a
given population ($\rho=\int^{10^{13}}_{10^5} \phi({\cal M}) d{\cal M}$).

The stellar mass density of quiescent galaxies (shown in Figure
\ref{LD_Ell}) increases by 1.1 dex (a factor of 14) between $z=1.5-2$
and $z=0.8-1$, and still increases by 0.2 dex between $z=0.8-1$ and
$z=0.2-0.4$ (by 0.3 dex if we consider the local measurement by Bell
et al. 2003). Therefore, the stellar mass assembly of quiescent
galaxies appears to slow down at $z<1$.

Figure \ref{LD_Ell} shows the stellar mass density of red elliptical
galaxies. The lower and upper limits of the shaded areas correspond to
the C09 and G-C morphological classifications, respectively. At
$z<0.8$, the total stellar mass of red elliptical galaxies contributes
more than 80\% (70\%) of the total stellar mass in quiescent
galaxies.\\

\begin{figure*}[htb!]
\includegraphics[width=17.3cm,height=17.3cm]{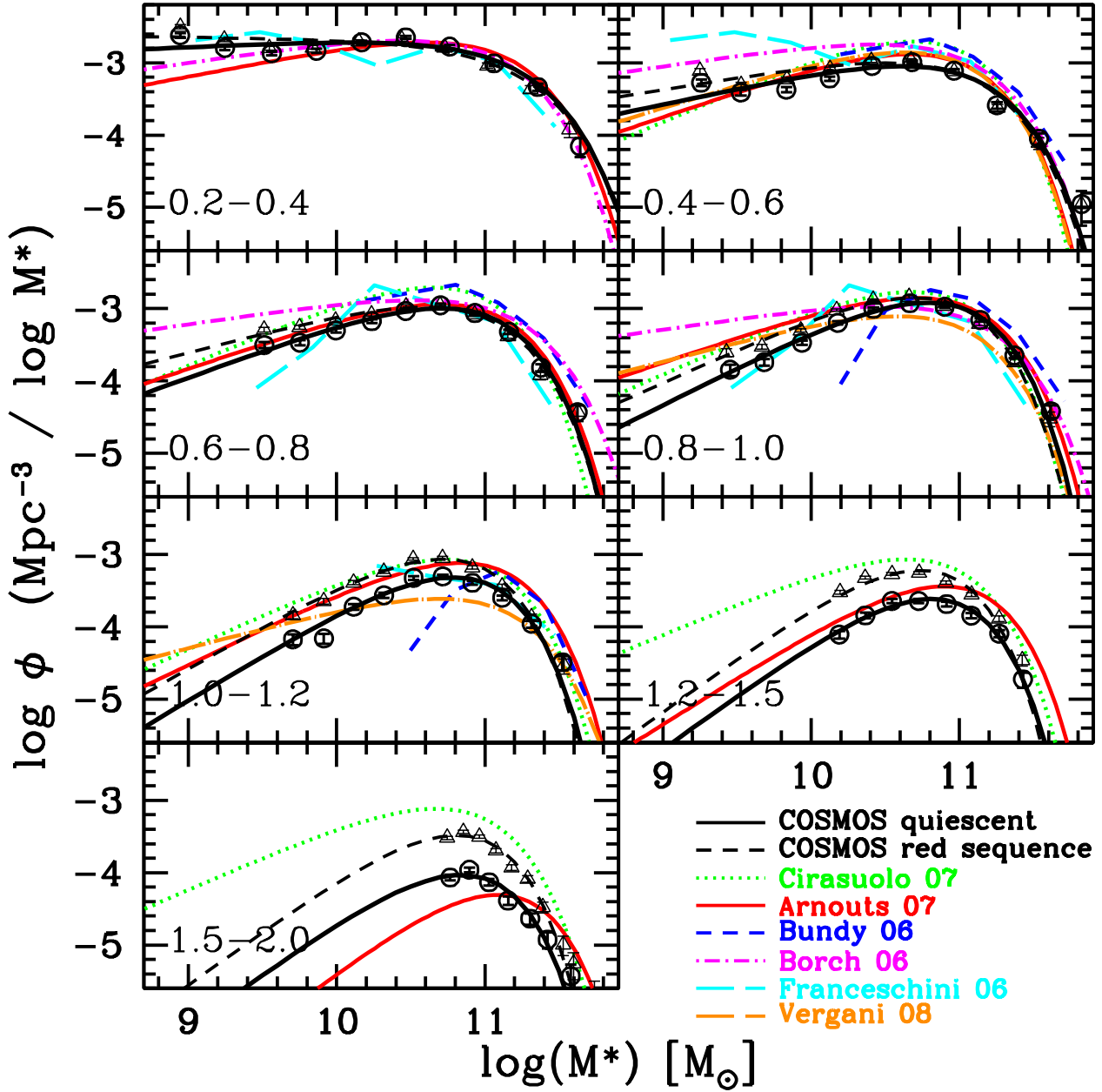}
\caption{Comparison between our MFs derived for quiescent (black open
  circles) and red sequence (open triangles) galaxies and the MFs from
  the literature. The MFs measured by Arnouts et al. (2007) (solid red
  curves) are based on a template-fitting classification. The MFs
  measured by Cirasuolo et al. (2007) (green dotted lines) are
  obtained using a red sequence classification. \label{MF_red_litt}}
\end{figure*}

\begin{figure*}[htb!]
\includegraphics[width=17.cm,height=16.8cm]{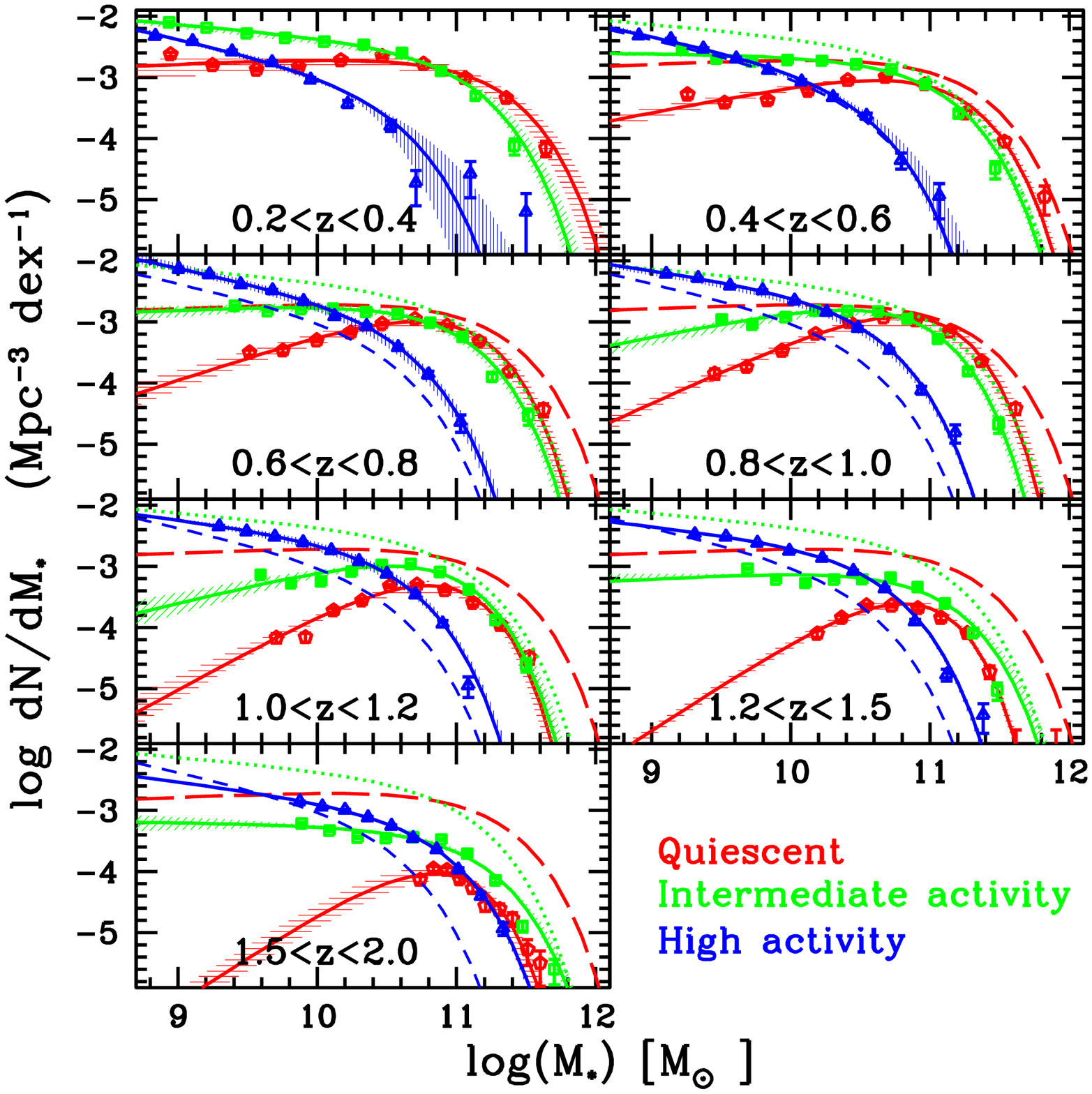}
\caption{MF by spectral type. The sample is split into ``high
  activity'' (blue vertical shaded area), ``intermediate activity''
  (green oblique shaded area) and ``quiescent'' (red horizontal shaded
  area) galaxies. The lower and upper limits of the shaded areas are
  the extreme values of the MFs estimated in four quadrants of
  0.5-deg$^2$, which quantifies the impact of cosmic variance. The
  blue short-dashed lines, the green dotted lines and the red
  long-dashed lines are the MFs measured at $z=0.2-0.4$ for the ``high
  activity'', ``intermediate activity'' and ``quiescent'' galaxies,
  respectively. \label{MF_blue_red}}
\end{figure*}

\subsection{Comparison with the literature}\label{litt}

Figure \ref{MF_red_litt} compares our ``quiescent'' MFs with those
from the literature. The K-band luminosity functions from Cirasuolo et
al. (2007) and Arnouts et al. (2007) are converted into stellar mass
functions using the mass-to-light ratio given in Arnouts et
al. (2007).

Most of the MFs from the literature are derived for red sequence
galaxies. Therefore, we split the COSMOS sample into ``red sequence''
and ``blue cloud'' galaxies according to the empirical limit
$M_{NUV}-M_R=0.5\;log({\cal M}) - 0.8\;z - 0.5$ (see Figure
\ref{bimoMorpho}). Our MFs for red sequence galaxies (open triangles
in Figure \ref{MF_red_litt}) are systematically higher at $z>1$ than
our MFs for quiescent galaxies (open circles). We interpret this
difference as being due to the presence of dust-extincted spirals
within the red sequence (see also Figure \ref{bimoColor}). This
contamination of the red sequence could increase with redshift since
the density of dusty star-forming galaxies increases with redshift
(e.g. Takeuchi et al. 2007, Le Floc'h et al. 2005).

The agreement between the various surveys at the high-mass end is good
(within 0.2 dex), given the uncertainties in the stellar masses, the
different classification methods and cosmic variance. The increase of
massive red galaxies between $z=1.5-2$ and $z=0.8-1$ has also been
seen in previous studies (e.g. Cirasuolo et al. 2007, Arnouts et
al. 2007). However, the amplitude of this increase differs between the
samples. At $z=1.5-2$, the MF normalization derived by Cirasuolo et
al. (2007) is a factor $\sim$10 higher than Arnouts et al. (2007). Our
MFs for red sequence (open triangles) and quiescent galaxies (open
circles) show that the two measurements can be partly reconciled when
consistent selection criteria are applied (Cirasuolo et al.  used red
sequence galaxies and Arnouts et al. used a template fitting
classification close to our quiescent definition).

\begin{figure}[htb!]
\begin{tabular}{c}
\includegraphics[width=7.9cm]{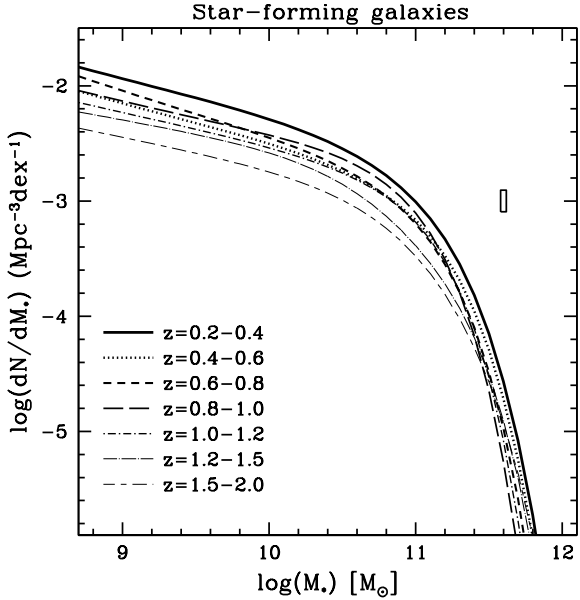}
\end{tabular}
\caption{MF of ``star-forming''
galaxies (sum of intermediate and high activity galaxies) from $z=2$
to $z=0.2$. The vertical box quantifies the cosmic variance at
$z=0.2-0.4$ (Scoville et al.  2007).
\label{MF_starforming}}
\end{figure}

\section{Stellar mass assembly of star-forming galaxies}\label{late}

This section presents the MF and stellar mass density for star-forming
galaxies. We subdivided the star-forming sample into ``intermediate
activity'' and ``high activity'' galaxies using a best-fit template
procedure (see \S\ref{template} and Figure
\ref{Figclassification}). We did not attempt to introduce a
morphological separation since it would require a too fine
classification within the Spi/Irr population.

\subsection{MF of star-forming galaxies}\label{MFlate}

Figure \ref{MF_blue_red} shows the MF evolution of the ``intermediate
activity'' (green oblique shaded area) and ``high activity'' galaxies
(blue vertical shaded area). The MF of the ``quiescent'' galaxies (red
horizontal shaded area, see \S\ref{MFearly}) is added as a reference.
Both the MFs of ``intermediate activity'' and ``high activity''
galaxies evolve between $z=2$ and $z=0.2$. Since their shapes change
with time, this evolution is mass-dependent.

As a consistent trend at all redshifts, the slope of the ``high
activity'' galaxies is always the steepest. The density of ``high
activity'' galaxies decreases with cosmic time but the size of this
decrease is a strong function of the stellar mass. Between $z=1.2-1.5$
and $z=0.2-0.4$, their number density decreases by a factor of 5 at
$log({\cal M})>11$, and only by a factor of 1.1 at $9.5<log({\cal
  M})<10$.

The ``intermediate activity'' MF follows a different
evolution. Galaxies as massive as $log({\cal M}) \sim 11.6$ are
already in place at $z=1.5-2$. The density of lower mass galaxies
rises with times.

We plotted in Figure \ref{MF_starforming} the MFs of all star-forming
galaxies, i.e. the sum of the ``intermediate activity'' and ``high
activity'' MFs. The exponential cutoff does not evolve by more than
0.2 dex between $z=1.2-1.5$ and $z=0.2-0.4$, which is consistent with
cosmic variance and systematic errors in stellar mass measurements. We
do not observe significant changes in the MF shape. Therefore, the
decrease with time of ``high activity'' galaxies is partly
counter-balanced by the build up of the ``intermediate activity'' MF.

\begin{figure}[htb!]
\includegraphics[width=7.9cm]{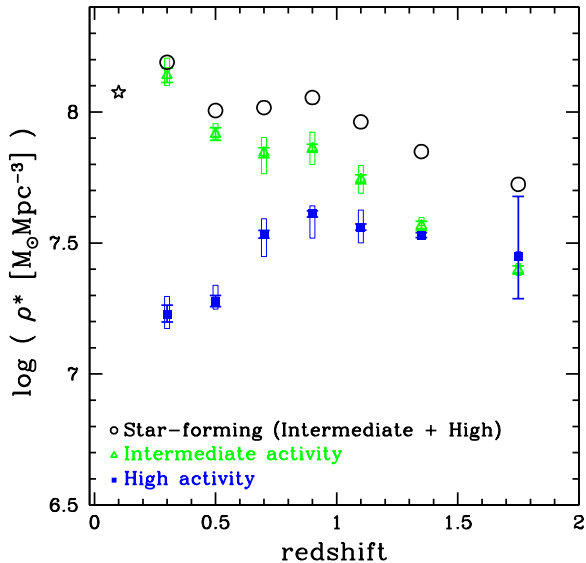}
\caption{Stellar mass density of star-forming galaxies.  The green
  open triangles and the blue solid squares are the ``intermediate
  activity'' and ``high activity'' galaxies, respectively. The black
  open circles are all star forming galaxies (sum of ``high activity''
  and ``intermediate activity'').  The open star is the local
  measurement by Bell et al. (2003).  The boxes correspond to the two
  extreme values measured in 4 quadrants of 0.5-deg$^2$. \label{LD}}
\end{figure}

\subsection{Stellar mass density of star-forming galaxies}\label{LDlate}

Figure \ref{LD} shows the evolution of the integrated stellar mass
density for ``intermediate activity'' (green triangles), ``high
activity'' (blue squares) and all star-forming galaxies (sum of
intermediate and high activity: open circles).

The stellar mass density increases between $z=1.5-2$ and $z=0.8-1$ for
star-forming populations. The density increases by 0.5 dex, 0.2 dex
and 0.3 dex for the ``intermediate activity'', the ``high activity''
and all star-forming galaxies, respectively.

The ``intermediate activity'' and ``high activity'' galaxies follow an
opposite evolution at $z<1$. Between $z=0.8-1$ and $z=0.2-0.4$, the
stellar mass density increases by 0.3 dex for the ``intermediate
activity'' galaxies, while it decreases by 0.4 dex for the ``high
activity'' galaxies. The stellar mass density of all star forming
galaxies shows little evolution at $z<1$. Indeed, the stellar mass
density at $z\sim 0.1$ measured by Bell et al. (2003) for star-forming
galaxies is consistent with the stellar mass density we measure at $z
\sim 1$.

\begin{figure}[htb!]
  \includegraphics[width=7.9cm]{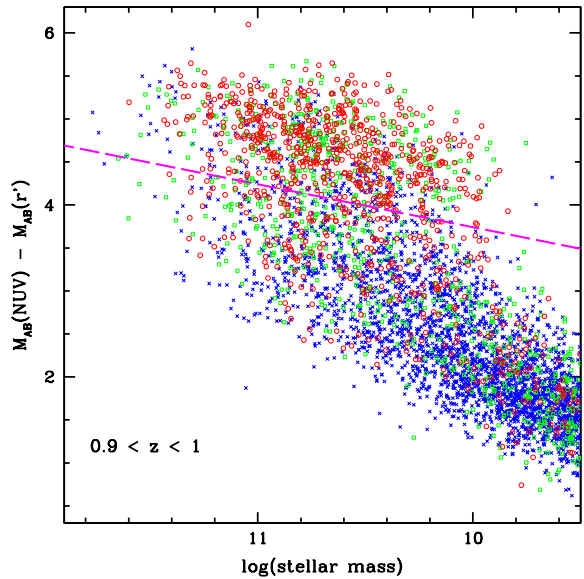}
\caption{Rest-frame color $ M({\rm NUV}) - M(r^+) $ (not corrected for
  dust reddening) as a function of stellar mass at $i^+<24$ in a given
  redshift bin $0.9<z<1$. The red open circles are the galaxies
  morphologically selected as elliptical by C09. The green open
  squares are selected as elliptical with the G-C parameters and not
  by C09. The blue crosses are Spi/Irr galaxies classified with the
  G-C parameters. The magenta dashed line is the limit adopted to
  split the sample into red sequence and blue cloud
  galaxies.\label{bimoMorpho}}
\end{figure}

\begin{figure*}[htb!]
\includegraphics[width=17.cm,height=16.cm]{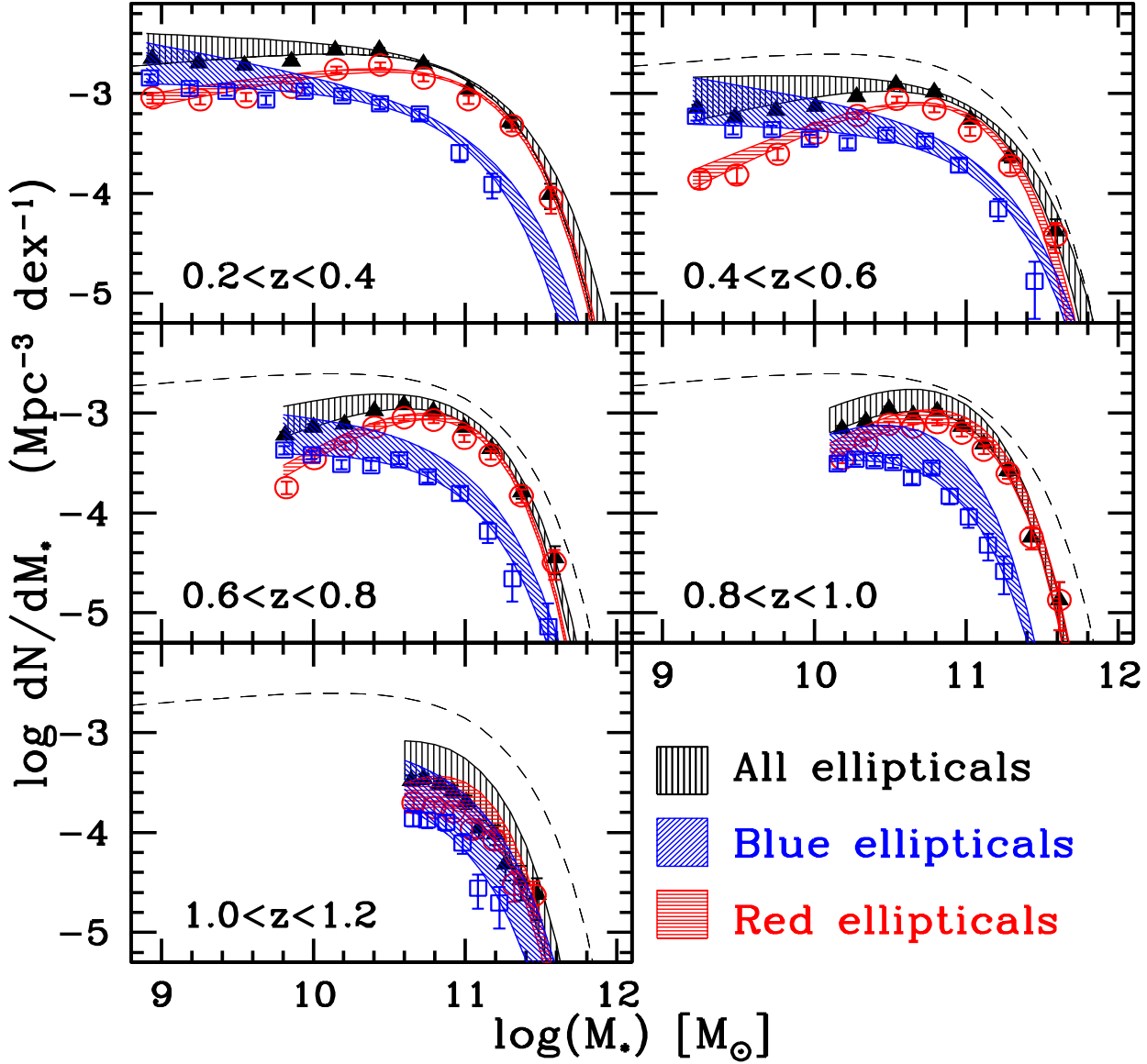}
\caption{MF of elliptical galaxies selected in morphology without any
  cut in color (black vertical shaded area). The blue oblique shaded
  areas and the horizontal red shaded areas are the MFs of the blue
  and red elliptical galaxies, respectively.  The upper and lower
  envelope MFs are obtained using the G-C and C09 morphological
  classifications, respectively. For clarity in the figure, the
  non-parametric estimates are shown only for the C09
  classification. The dashed line is the MF of elliptical galaxies
  derived at $z=0.2-0.4$, which is shown in each redshift bin to serve
  as a reference.\label{MF_TD}}
\end{figure*}

\begin{figure}[htb!]
\includegraphics[width=7.9cm]{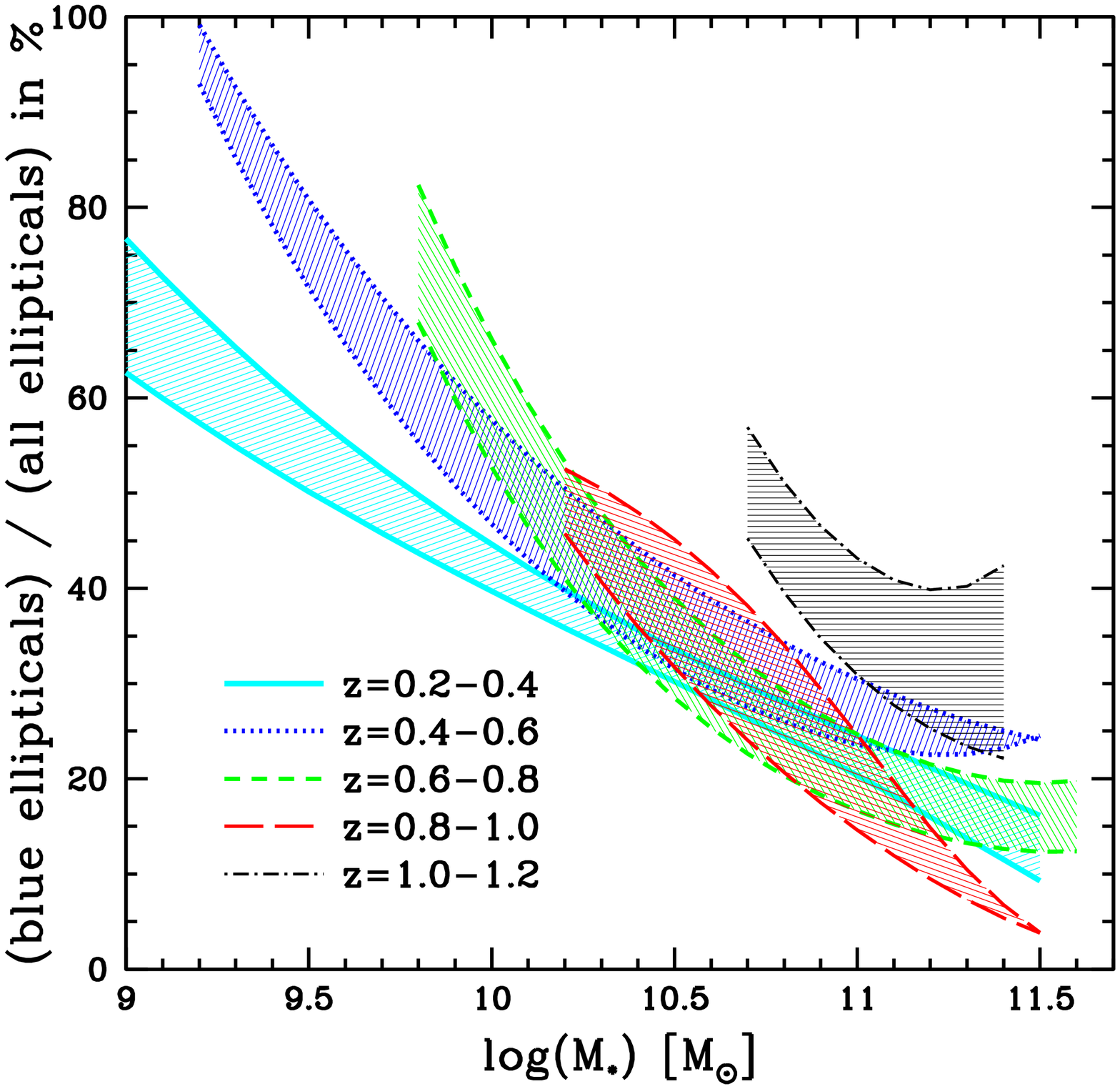}
\caption{Fraction (in \%) of blue galaxies in the elliptical sample
  (morphologically selected). The upper and lower limits are obtained
  using the G-C and C09 morphological classifications,
  respectively. \label{compositionTBB}}
\end{figure}

\section{Star-forming galaxies with an elliptical morphology}\label{blueEll}

As shown in Figure \ref{bimoMorpho}, the elliptical galaxies
preferentially have a red color. However, a significant population of
elliptical galaxies appears to be blue. The presence of ``blue
elliptical'' is not unexpected.  The ``blue elliptical'' galaxies
could be newly formed elliptical galaxies still consuming the gas of
their progenitors.  Accretion of new cold gas into an old elliptical
galaxy (Hammer et al. 2007) could also produce a blue color. We
present here the MF of this elliptical population which is
star-forming.

Figure \ref{MF_TD} shows the MFs of the elliptical galaxies purely
selected by morphology (black vertical shaded area), the ``blue
elliptical'' galaxies (blue shaded area) and the red elliptical
galaxies (red shaded area). The shape of the ``blue elliptical'' MFs
differs from those of the red elliptical galaxies: the slope is
steeper and the exponential cutoff is shifted to lower mass (in
agreement with Ilbert et al. 2006a). As a consequence of these
different shapes, the contribution of the ``blue elliptical'' galaxies
to the total elliptical population depends strongly on stellar
mass. Figure \ref{compositionTBB} shows that, regardless of redshift,
the fraction of ``blue ellipticals'' decreases towards high mass
systems.  The ``blue elliptical'' galaxies represent $<20\%$ of the
massive elliptical galaxies (at $log({\cal M})>11$ and $z<1$), but
their contribution reaches 40-60\% at $log({\cal M}) \sim 10$.

\begin{figure}[htb!]
\includegraphics[height=7.7cm]{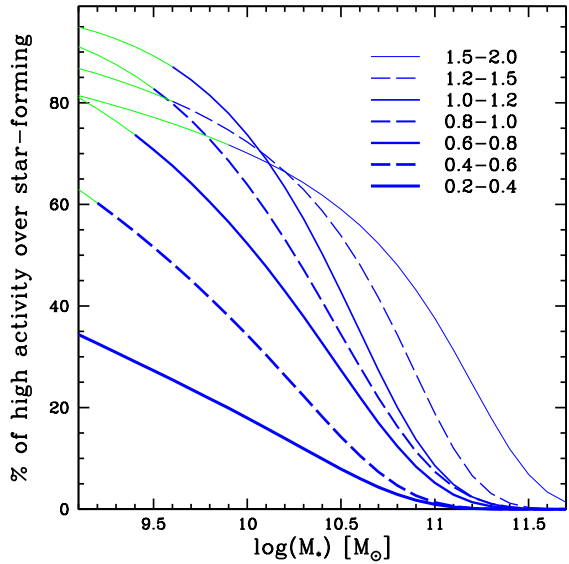}
\caption{Fraction in \% of ``high activity'' galaxies within the
  star-forming sample as a function of the stellar mass and per
  redshift bin (MF of the ``high activity'' divided by the MF of all
  star-forming galaxies). The thick blue line corresponds to the
  stellar mass range covered by the data. The thin green line is
  obtained using the extrapolation of the MF with a Schechter
  function. \label{fractionSF}}
\end{figure}

\section{Discussion}\label{discussion}

\subsection{Redistribution of the star formation activity along cosmic time}\label{massiveSF}

We first discuss our results on the MF of star forming galaxies.

The total stellar mass of ``high activity'' galaxies decreases by a
factor of 1.7 between $z=1.5-2$ and $z=0.2-0.4$ (see \S\ref{late}),
while the stellar mass of a given star-forming galaxy grows. A simple
interpretation is that, after an intense star formation activity
period, these galaxies evolve into less active systems (intermediate
activity or quiescent). Based on the BC03 models, a passive evolution
could transform a ``high activity'' galaxy at $z\sim 1.5$ into an
``intermediate activity'' galaxy at $z\sim 0.7$ \footnote{A galaxy
  with a rest-frame color $( {\rm NUV}-r^+)_{template}\sim 1$ reddens
  by 2 mag in 3 Gyr assuming an exponentially decreasing SFR with
  $\tau=1$ Gyr and a solar metallicity.}.

Figure \ref{fractionSF} shows the fraction of ``high activity''
galaxies within the star-forming sample. This fraction does not
decrease uniformly at all stellar masses. Between $z=1.5-2$ and
$z=1-1.2$, the fraction of ``high activity'' galaxies at $log({\cal
  M})\sim 11$ drops by a factor of $\sim$4 (from 40\% to 10\%), while
the fraction stays approximately constant at low mass $log({\cal
  M})\sim 9.5-10$. The contribution of low mass, high activity
galaxies starts to decrease significantly only at $z<1$. Since each
spectral type corresponds to a range of SSFR (see \S\ref{template}),
it implies that the low mass star-forming galaxies are able to
maintain a high SSFR, while the massive galaxies evolve rapidly into
systems with a lower SSFR. Therefore, the redistribution of the star
formation activity follows a clear ``downsizing'' pattern (Cowie et
al. 1996) within the star-forming sample itself.

The MF of all star-forming galaxies (sum of intermediate and high
activity) shows little evolution between $z\sim 1.5$ and $z\sim 0.2$.
The re-distribution of the star formation activity between
``intermediate activity'' and ``high activity'' galaxies does not
affect the overall mass distribution of the star-forming galaxies. The
little evolution of the star-forming MF means that a fraction of
star-forming galaxies is transferred to the quiescent population (as
already noted by Arnouts et al. 2007 and Cowie et al. 2008), since
star forming galaxies generate new stellar populations between $z\sim
1.5$ and $z\sim 0.2$. Using backward evolution models, Boissier et
al. (2009) discuss in detail the consistency between the little
evolution of the star-forming MF and the building of the quiescent
population. We discuss in the next sections which processes can
generate these quiescent systems.

\subsection{Mass-dependent assembly of elliptical galaxies at $z<1$}\label{evolution}

We find that the most massive quiescent galaxies are already in place
at $z\sim 1$ while their density still rises at low/intermediate
masses at $z<1$ (see \S\ref{MFearly}). This mass-dependent evolution
of quiescent galaxies confirms the ``downsizing'' pattern found by the
COMBO-17 survey (Borch et al. 2006) and the DEEP2 survey (Bundy et
al. 2006, Cimatti et 2006). We also investigated the MF evolution of
the quiescent galaxies with an elliptical morphology. We retrieved the
same downsizing pattern: the most massive red elliptical galaxies are
already in place at $z\sim 1$, while the low/intermediate mass red
E/S0 galaxies are still being created at $z<1$. In principle, the weak
evolution of the massive red elliptical galaxies at $z<1$ could be
explained by a selection procedure based on the galaxy spectra (Van
Dokkum \& Franx 2001) since blue elliptical galaxies missed by a
multi-color criterion could account for a significant evolution of the
high-mass end. However, we showed in section \S\ref{MFearly} that the
contribution of blue elliptical galaxies is limited at 20\% at
$log({\cal M})>11$. Therefore, the blue elliptical contribution can
not explain the low evolution rate of the most massive elliptical
galaxies at $z<1$.

A possible interpretation of this mass dependent evolution is a galaxy
assembly by merger process more efficient at low/intermediate mass
than at high mass at $z<1$, if we assume that red elliptical galaxies
are formed by mergers (e.g. Toomre \& Toomre 1972, Athanassoula 2008,
Bekki et al. 2008)\footnote{Bekki et al. (2008) simulated merging
  between gas rich spiral galaxies of mass ${\cal M} \sim 10^{9} {\cal
    M}_\Sun$, which formed a dynamically relaxed low mass elliptical
  galaxy in 1.4 Gyr with a SFR of 0.03 $M_\Sun/yr$.}. This picture is
in agreement with De Ravel et al. (2008) who found that the merger
rate decreases with stellar mass when using galaxy pair counts in the
VIMOS-VLT Deep Survey (le F\`evre et al. 2005).

The formation of low mass quiescent galaxies with an elliptical
morphology could also be explained by ``morphological quenching''
(Martig et al. 2009). The presence of a spheroid is sufficient to
stabilize the gas disk and quench the star formation. This process is
efficient in low mass halos which could explain the formation of these
low mass, quiescent and elliptical galaxies.

\begin{figure}[htb!]
\includegraphics[width=7.9cm]{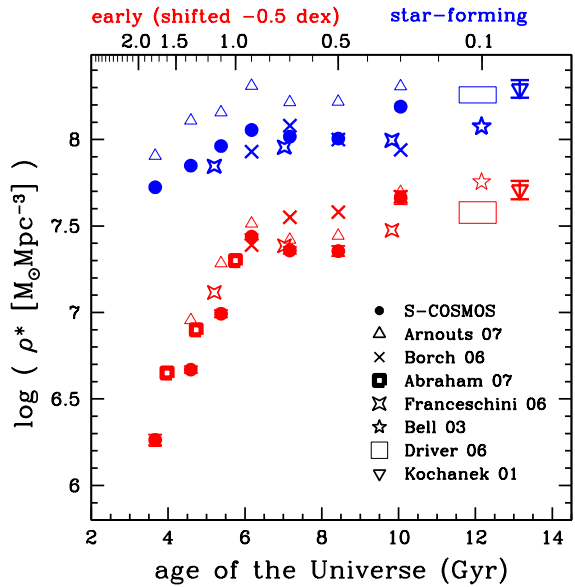}
\caption{Evolution of the stellar mass density of star-forming (blue filled circles) and quiescent galaxies (red filled circles) compared to various results from the literature. The stellar mass density of all the red sequence and quiescent galaxies is shifted vertically by -0.5 dex for the clarity of the figure. The offset of 0.2 dex between Arnouts et al. (2007) and our measurement for star-forming galaxies is discussed in appendix \ref{masstolight}.\label{LD_Litt}}
\end{figure}

\begin{figure}[htb!]
\includegraphics[height=7.9cm]{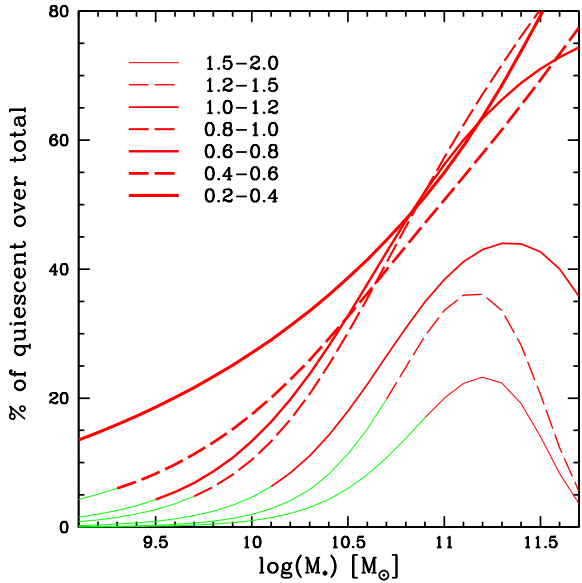}
\caption{Fraction of the ``quiescent'' over the total population per stellar mass bin (MF of the quiescent divided by the total MF). The thick red line corresponds to the stellar mass range covered by the data. The thin green line is obtained using the extrapolation of the MF with the Schechter function.   \label{fractionEll}}
\end{figure}

\subsection{Slow down in the assembly of massive quiescent galaxies at $z \sim 1$}\label{evolution}

The most massive quiescent galaxies are not in place at $z \sim 2$ and
their number density evolves rapidly between $z\sim 2$ and $z\sim 1$.
The exponential cutoff of their MF increases by 0.4 dex between
$z=1.5-2$ and $z=0.8-1$. Figure \ref{LD_Litt} shows that the rapid
assembly rate of quiescent galaxies at $1<z<2$ is consistent with
other surveys, despite the different methods used to classify early
type galaxies, to determine the stellar masses and to measure the
distances (spectro-z or photo-z). Using deeper data in the GOODS
  field, we also checked that our results don't suffer from a
  significant incompleteness which could mimic the rapid assembly of
  quiescent galaxies at $1<z<2$ (see appendix \ref{GOODS}).

Therefore, the most massive quiescent galaxies are created rapidly
between $z\sim 2$ and $z\sim 1$ and their assembly slows down at
$z<1$. We tentatively explain this slow down in their evolution by
analyzing the relative evolution of the quiescent and star-forming MFs
from $z=2$ to $z=0.2$.

Figure \ref{fractionEll} shows the fraction of quiescent galaxies as a
function of stellar mass. The quiescent galaxies represent less than
20\% of the most massive galaxies at $z=1.5-2$. As a consequence,
``wet mergers" between massive star-forming galaxies directly create
new massive quiescent galaxies (since star-forming galaxies are more
massive and more numerous) which generates a rapid growth of the
quiescent high mass end between $z\sim 2$ and $z\sim 1$.

A change of regime occurs at $z<1$, where the exponential cut-off of
star-forming galaxies is shifted at lower mass than the exponential
cut-off of quiescent galaxies (see Figure \ref{MF_blue_red}). This
change is likely explained by the rapid increase of the massive
quiescent population at $1<z<2$, combined with the decrease of the SFR
at $z<1$ (e.g. Lilly et al. 1996, Le Floc'h et al. 2005, Tresse et
al. 2007) which prevents to regenerate the massive star-forming
population. Therefore, the quiescent population dominates the massive
end of the MF at $z<1$, with more than 60-70\% of quiescent galaxies
at $log({\cal M})>11-11.5$.  As a consequence, ``wet mergers" become
inefficient at $z<1$ to generate the most massive quiescent galaxies.

Dry merging (merging between quiescent galaxies, e.g. Van Dokkum et
al. 2001) is the only process left to form the most massive quiescent
galaxies at $z<1$. This process has less impact on the quiescent MF
evolution since ``dry merging'' is not a direct supply of new
quiescent systems (the progenitors are already quiescent
galaxies). Moreover, ``dry merging'' involving two massive quiescent
galaxies ($log({\cal M})>11$) is not a common process since the
exponential cutoff of the quiescent MF does not evolve significantly
at $z<1$.

Therefore, the disappearance of ``wet merging" as an efficient process
to form the most massive ellipticals could explain the slow down in
their assembly at $z<1$.

\begin{figure}[htb!]
\includegraphics[width=7.7cm]{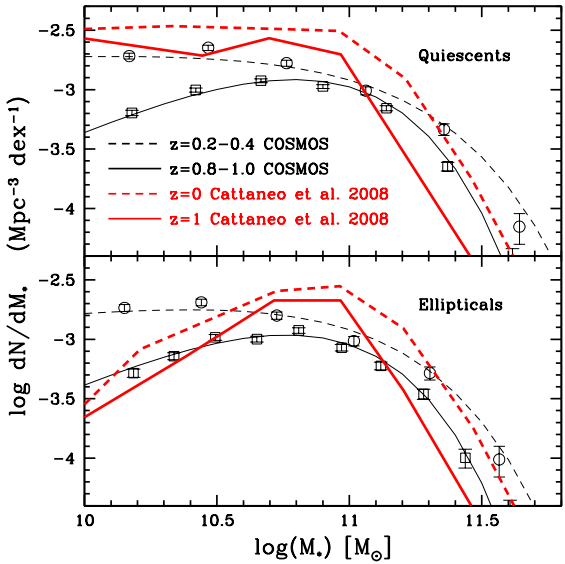}
\caption{MF of ``quiescent'' galaxies (thick red lines in the top panel) and elliptical galaxies (thick red lines in the bottom panel) at $z=0$ and $z=1$ predicted by Cattaneo et al. (2008) using the GalICS semi-analytical simulations and a ``halo quenching'' model. Our quiescent and red elliptical MFs are reported in each panel. \label{MF_simul}}
\end{figure}

\subsection{Indirect constraints on the AGN feedback}

Two different modes of AGN activity are usually considered: the
``bright mode'' (e.g. QSO) and the ``radio mode'' which radiates less
energy than the ``bright mode'' but is a more common mechanism
(e.g. Croton et al. 2006). In both modes, the AGN feedback can prevent
the gas to cool in the dark matter halo. Since the ``radio mode'' is
less energetic, the radio jet from the black hole prevent the gas to
cool only in quasi-hydrostatic shock-heated halos, i.e. in halos above
a critical mass around $10^{12}M_{\Sun}$ (Cattaneo et al. 2006,
Somerville et al. 2008).

As a consequence, the radio mode prevents the star formation of
recurring in all the galaxies of massive halos. Therefore, the star
formation is shut down in the galaxy hosting the AGN, but also in all
the galaxies surrounding the AGN. We discuss here the impact of AGN
feedback on the ``surrounding galaxies''\footnote{We call
  ``surrounding galaxies'' the galaxies which belong to the same halo
  as the AGN but which are not hosting the AGN themselves. These
  galaxies are satellite galaxies if the AGN is hosted by the central
  galaxy of the dark matter halo.}. A specific study on the host
galaxies of the radio sources is presented in Smol{\v c}i{\'c} et
al. (2009).

The shutdown of the star formation in the ``surrounding galaxies''
does not modify their morphology. As their original morphology can be
any, we then expect a significant population of quiescent galaxies
with a Spi/Irr morphology. Indeed, the ratio between the red
elliptical and the quiescent MFs shows a significant fraction of
Spi/Irr having a quenched star formation at low masses (40-60\% at
$log({\cal M}) \sim 9.5$, see Figure \ref{compositionMF}), which
leaves room for this ``external'' action mode of AGN feedback.

More quantitatively, Cattaneo et al. (2008) provided the predicted MFs
for quiescent (red sequence in Cattaneo et al.) and elliptical
(bulge-dominated in Cattaneo et al. ) galaxies at $z=0$ and
$z=1$. They used a ``halo quenching'' model (Somerville et
al. 2008). In this model, the AGN feedback shuts down the star
formation in all the galaxies within a halo more massive than
$10^{12}M_{\Sun}$ (Cattaneo et al. 2006, Somerville et al. 2008). We
report in Figure \ref{MF_simul} the MF predicted by Cattaneo et
al. (2008). As expected, the predicted density of quiescent galaxies
stays constant at low/intermediate masses (top panel), while the
density of elliptical galaxies decreases toward low masses (bottom
panel). On the basis of these predicted MFs, $\sim 10\%$ of the
quiescent galaxies would be elliptical at $log({\cal M})\sim 10$. By
contrast, we found a fraction of 80\%(60\%) quiescent galaxies with an
elliptical morphology (see Figure \ref{compositionMF}). Therefore,
quenched star formation is more often linked to an elliptical
morphology than would be predicted by a ``halo quenching''
model. These measurements constrain in the same way all the processes
quenching star formation without modifying the galaxy morphology, for
example, gas starvation (i.e. satellite galaxies not fueled in cold
gas since they are not at the center of the dark matter halo potential
well).

A more quantitative constraint on these mechanisms requires a detailed
comparison with predictions of semi-analytical simulations, which is
planned for forthcoming papers.

\section{Conclusion}\label{Conclusion}

We derived the galaxy stellar mass function and stellar mass density
in the 2-deg$^2$ COSMOS field. We explored stellar mass assembly by
morphological and spectral type from $z=2$ to $z=0.2$. The MF estimate
is based on 196,000 galaxies selected at $F_{3.6\mu m} > 1 \mu Jy$ and
photo-z with an accuracy of $\sigma_{(z_{phot}-z_{spec})/(1+z_{spec})}
= 0.008$ at $i^+_{AB}>22.5$.  We summarize our results below.

$\bullet$ We found that $z\sim 1$ is an epoch of transition in the
assembly of quiescent galaxies. Their stellar mass density increases
by 1.1 dex between $z=1.5-2$ and $z=0.8-1$ (corresponding to a period
of 2.5 Gyr), but only by 0.3 dex between $z=0.8-1$ and $z \sim 0.1$
(a period of 6 Gyr). The high-mass exponential cutoff of the quiescent
MF increases by 0.4 dex between $z=1.5-2$ and $z=0.8-1$, but almost no
evolution is seen at $z<1$. We investigated if the weak evolution of
the most massive quiescent galaxies is also seen using a morphological
classification. The exponential cutoff of massive red elliptical
galaxies does not increase significantly at $z<1$. Moreover, the blue
elliptical galaxies do not contribute more than 20\% to the high-mass
end of the total elliptical sample, which is not sufficient to produce
significant evolution in the exponential cutoff.

$\bullet$ We found that the high-mass end of the star-forming MF is
shifted below the high-mass end of the quiescent MF at
$z<1$. Therefore, we interpreted the slow down in the assembly of the
most massive elliptical galaxies at $z<1$ as being due to a ``lack of
supply'' of massive star-forming galaxies available for ``wet
mergers".

$\bullet$ We observed a rapid rise of quiescent galaxies at
low/intermediate masses. We characterized the nature of these newly
formed quiescent galaxies by adding morphological information. We
quantified the fraction of quiescent galaxies with an elliptical
morphology, as well as the fraction of Spi/Irr galaxies with quenched
star formation. The significant fraction of quenched Spi/Irr
($40-60\%$ at $log({\cal M}) \sim 9.5$) leaves room for a mechanism
which shuts down the star formation without transforming their
morphology, such as the impact of AGN feedback on the satellite
galaxies of a massive halo (e.g. Cattaneo et al. 2006). Since the
majority of quiescent galaxies have an elliptical morphology at
$z<0.8$ (80-90\% at $log({\cal M})\sim 11$), the dominant process
which shuts down star formation should be linked to the acquisition of
an elliptical morphology, as might be expected in galaxy merging  and/or morphological quenching (Martig et al. 2009).

$\bullet$ Finally, we divided the star-forming sample into
``intermediate activity'' and ``high activity'' galaxies, which
corresponds to two classes of SSFR (SFR divided by stellar mass). The
MF of the ``high activity'' galaxies shows that the most massive of
them end their high activity phase first. Therefore, the low mass
star-forming galaxies are able to maintain a high SSFR, while the
massive galaxies evolve rapidly into systems with a lower SSFR. This
redistribution of the star formation activity follows a clear
``downsizing'' pattern (Cowie et al. 1996) within the star-forming
sample itself.

\acknowledgments

We are grateful to the referee for his/her careful reading of the
manuscript and his/her useful suggestions.

This work is based on observations made with the Spitzer Space
Telescope, which is operated by the Jet Propulsion Laboratory,
California Institute of Technology under NASA contract 1407.  Support
for this work was provided by NASA through Contract Number 1278386
issued by JPL. We gratefully acknowledge the contributions of the
entire COSMOS collaboration consisting of more than 100 scientists.
The {\it HST} COSMOS program was supported through NASA grant
HST-GO-09822.  More information on the COSMOS survey is available at
http://www.astro.caltech.edu/cosmos.

\clearpage

\begin{table*}[htb!]
\begin{center}
\begin{tabular}{l c c c c c c c} \hline

     &       &         &                       &         & \multicolumn{1}{c}{$log({\cal M}^*$}) &                  \multicolumn{1}{c}{$\Phi^*$} & $log(\rho^*)$  \\ 
type & z-bin &  Number & $log({\cal M}_{low})$ &  $\alpha$ &         (${\cal M}_\Sun$)             &             ($10^{-3} Mpc^{-3}$)        & (${\cal M}_\Sun Mpc^{-3}$)         \vspace{0.2cm} \\ \hline
Quiescent&          0.2-0.4 &  2202 &   8.8 & -0.91$^{{\rm + 0.02}}_{{\rm - 0.02}}$ & 11.13$^{{\rm + 0.03}}_{{\rm - 0.03}}$ &  1.12$^{{\rm + 0.07}}_{{\rm - 0.07}}$ &  8.16$^{{\rm + 0.04}}_{{\rm - 0.04}}$ \\ 
&                   0.4-0.6 &  1708 &   9.1 & -0.56$^{{\rm + 0.03}}_{{\rm - 0.03}}$ & 10.97$^{{\rm + 0.03}}_{{\rm - 0.03}}$ &  0.87$^{{\rm + 0.04}}_{{\rm - 0.04}}$ &  7.85$^{{\rm + 0.03}}_{{\rm - 0.03}}$ \\ 
     &              0.6-0.8 &  2432 &   9.4 & -0.25$^{{\rm + 0.04}}_{{\rm - 0.04}}$ & 10.83$^{{\rm + 0.02}}_{{\rm - 0.02}}$ &  1.15$^{{\rm + 0.03}}_{{\rm - 0.03}}$ &  7.86$^{{\rm + 0.02}}_{{\rm - 0.02}}$ \\ 
     &              0.8-1.0 &  3381 &   9.3 &  0.04$^{{\rm + 0.03}}_{{\rm - 0.03}}$ & 10.77$^{{\rm + 0.01}}_{{\rm - 0.01}}$ &  1.43$^{{\rm + 0.03}}_{{\rm - 0.03}}$ &  7.94$^{{\rm + 0.02}}_{{\rm - 0.02}}$ \\ 
     &              1.0-1.2 &  1447 &   9.6 &  0.25$^{{\rm + 0.08}}_{{\rm - 0.08}}$ & 10.70$^{{\rm + 0.03}}_{{\rm - 0.02}}$ &  0.55$^{{\rm + 0.02}}_{{\rm - 0.02}}$ &  7.49$^{{\rm + 0.02}}_{{\rm - 0.02}}$ \\ 
     &              1.2-1.5 &  1069 &  10.1 &  0.50                            & 10.64$^{{\rm + 0.14}}_{{\rm - 0.13}}$ &  0.26$^{{\rm + 0.02}}_{{\rm - 0.02}}$ &  7.17$^{{\rm + 0.02}}_{{\rm - 0.02}}$ \\ 
     &              1.5-2.0 &   468 &  10.7 &  0.50                            & 10.67$^{{\rm + 0.10}}_{{\rm - 0.10}}$ &  0.10$^{{\rm + 0.02}}_{{\rm - 0.04}}$ &  6.78$^{{\rm + 0.03}}_{{\rm - 0.03}}$ \\ 
\hline
Red sequence&       0.2-0.4 &  2343 &   8.8 & -1.03$^{{\rm + 0.04}}_{{\rm - 0.04}}$ & 11.18$^{{\rm + 0.05}}_{{\rm - 0.05}}$ &  0.86$^{{\rm + 0.14}}_{{\rm - 0.14}}$ &  8.12$^{{\rm + 0.05}}_{{\rm - 0.05}}$ \\ 
&                   0.4-0.6 &  2000 &   9.1 & -0.66$^{{\rm + 0.03}}_{{\rm - 0.03}}$ & 10.97$^{{\rm + 0.03}}_{{\rm - 0.03}}$ &  0.88$^{{\rm + 0.05}}_{{\rm - 0.05}}$ &  7.86$^{{\rm + 0.03}}_{{\rm - 0.03}}$ \\ 
     &              0.6-0.8 &  2752 &   9.4 & -0.46$^{{\rm + 0.03}}_{{\rm - 0.03}}$ & 10.86$^{{\rm + 0.02}}_{{\rm - 0.02}}$ &  1.09$^{{\rm + 0.04}}_{{\rm - 0.04}}$ &  7.84$^{{\rm + 0.02}}_{{\rm - 0.02}}$ \\ 
     &              0.8-1.0 &  4108 &   9.3 & -0.07$^{{\rm + 0.03}}_{{\rm - 0.03}}$ & 10.73$^{{\rm + 0.01}}_{{\rm - 0.01}}$ &  1.66$^{{\rm + 0.04}}_{{\rm - 0.04}}$ &  7.94$^{{\rm + 0.02}}_{{\rm - 0.02}}$ \\ 
     &              1.0-1.2 &  2624 &   9.6 &  0.18$^{{\rm + 0.06}}_{{\rm - 0.06}}$ & 10.65$^{{\rm + 0.02}}_{{\rm - 0.02}}$ &  1.00$^{{\rm + 0.03}}_{{\rm - 0.03}}$ &  7.69$^{{\rm + 0.02}}_{{\rm - 0.02}}$ \\ 
     &              1.2-1.5 &  2568 &  10.1 &  0.50                            & 10.56$^{{\rm + 0.013}}_{{\rm - 0.11}}$ &  0.64$^{{\rm + 0.03}}_{{\rm - 0.12}}$ &  7.49$^{{\rm + 0.02}}_{{\rm - 0.02}}$ \\ 
     &              1.5-2.0 &  1545 &  10.7 &  0.50                            & 10.63$^{{\rm + 0.09}}_{{\rm - 0.09}}$  &  0.35$^{{\rm + 0.05}}_{{\rm - 0.08}}$ &  7.30$^{{\rm + 0.03}}_{{\rm - 0.03}}$ \\ 
\hline
Red&                0.2-0.4 &  1396 &   8.8 & -0.76$^{{\rm + 0.03}}_{{\rm - 0.03}}$ & 11.02$^{{\rm + 0.03}}_{{\rm - 0.03}}$ &  1.37$^{{\rm + 0.09}}_{{\rm - 0.09}}$ &  8.12$^{{\rm + 0.04}}_{{\rm - 0.04}}$ \\ 
elliptical&         0.4-0.6 &  1020 &   9.1 & -0.35$^{{\rm + 0.04}}_{{\rm - 0.04}}$ & 10.86$^{{\rm + 0.03}}_{{\rm - 0.03}}$ &  0.90$^{{\rm + 0.04}}_{{\rm - 0.04}}$ &  7.77$^{{\rm + 0.03}}_{{\rm - 0.03}}$ \\ 
     &              0.6-0.8 &  1538 &   9.7 & -0.04$^{{\rm + 0.06}}_{{\rm - 0.06}}$ & 10.75$^{{\rm + 0.02}}_{{\rm - 0.02}}$ &  1.18$^{{\rm + 0.03}}_{{\rm - 0.03}}$ &  7.81$^{{\rm + 0.02}}_{{\rm - 0.02}}$ \\ 
     &              0.8-1.0 &  1902 &  10.1 &  0.04$^{{\rm + 0.08}}_{{\rm - 0.08}}$ & 10.75$^{{\rm + 0.03}}_{{\rm - 0.02}}$ &  1.28$^{{\rm + 0.03}}_{{\rm - 0.03}}$ &  7.86$^{{\rm + 0.02}}_{{\rm - 0.02}}$ \\ 
     &              1.0-1.2 &   480 &  10.6 &  0.50                            & 10.65$^{{\rm + 0.09}}_{{\rm - 0.09}}$ &  0.39$^{{\rm + 0.05}}_{{\rm - 0.05}}$ &  7.36$^{{\rm + 0.04}}_{{\rm - 0.05}}$ \\ 
\hline
Blue&               0.2-0.4 &  1484 &   8.8 & -1.29$^{{\rm + 0.03}}_{{\rm - 0.03}}$ & 11.10$^{{\rm + 0.09}}_{{\rm - 0.08}}$ &  0.32$^{{\rm + 0.05}}_{{\rm - 0.05}}$ &  7.71$^{{\rm + 0.06}}_{{\rm - 0.09}}$ \\ 
elliptical&         0.4-0.6 &  1474 &   9.1 & -1.24$^{{\rm + 0.03}}_{{\rm - 0.03}}$ & 11.09$^{{\rm + 0.07}}_{{\rm - 0.06}}$ &  0.22$^{{\rm + 0.03}}_{{\rm - 0.03}}$ &  7.52$^{{\rm + 0.05}}_{{\rm - 0.06}}$ \\ 
     &              0.6-0.8 &  1306 &   9.7 & -1.10$^{{\rm + 0.06}}_{{\rm - 0.06}}$ & 10.93$^{{\rm + 0.05}}_{{\rm - 0.05}}$ &  0.35$^{{\rm + 0.05}}_{{\rm - 0.05}}$ &  7.49$^{{\rm + 0.03}}_{{\rm - 0.03}}$ \\ 
     &              0.8-1.0 &  1160 &  10.1 & -0.23$^{{\rm + 0.13}}_{{\rm - 0.12}}$ & 10.58$^{{\rm + 0.04}}_{{\rm - 0.04}}$ &  0.87$^{{\rm + 0.04}}_{{\rm - 0.04}}$ &  7.48$^{{\rm + 0.02}}_{{\rm - 0.02}}$ \\ 
     &              1.0-1.2 &   468 &  10.6 & -1.00                            & 10.87$^{{\rm + 0.13}}_{{\rm - 0.14}}$ &  0.39$^{{\rm + 0.07}}_{{\rm - 0.07}}$ &  7.47$^{{\rm + 0.18}}_{{\rm - 0.19}}$ \\ 
\hline
Elliptical&         0.2-0.4 &  2880 &   8.8 & -1.06$^{{\rm + 0.02}}_{{\rm - 0.02}}$ & 11.13$^{{\rm + 0.04}}_{{\rm - 0.03}}$ &  1.28$^{{\rm + 0.09}}_{{\rm - 0.09}}$ &  8.26$^{{\rm + 0.04}}_{{\rm - 0.04}}$ \\ 
&                   0.4-0.6 &  2494 &   9.1 & -0.95$^{{\rm + 0.02}}_{{\rm - 0.02}}$ & 11.06$^{{\rm + 0.03}}_{{\rm - 0.03}}$ &  0.81$^{{\rm + 0.05}}_{{\rm - 0.05}}$ &  7.96$^{{\rm + 0.03}}_{{\rm - 0.03}}$ \\ 
     &              0.6-0.8 &  2844 &   9.7 & -0.61$^{{\rm + 0.04}}_{{\rm - 0.04}}$ & 10.87$^{{\rm + 0.02}}_{{\rm - 0.02}}$ &  1.44$^{{\rm + 0.07}}_{{\rm - 0.07}}$ &  7.98$^{{\rm + 0.02}}_{{\rm - 0.02}}$ \\ 
     &              0.8-1.0 &  3062 &  10.1 & -0.20$^{{\rm + 0.06}}_{{\rm - 0.06}}$ & 10.74$^{{\rm + 0.02}}_{{\rm - 0.02}}$ &  2.00$^{{\rm + 0.05}}_{{\rm - 0.06}}$ &  8.01$^{{\rm + 0.01}}_{{\rm - 0.01}}$ \\ 
     &              1.0-1.2 &   948 &  10.6 & -0.30                            & 10.76$^{{\rm + 0.13}}_{{\rm - 0.10}}$ &  0.93$^{{\rm + 0.06}}_{{\rm - 0.12}}$ &  7.68$^{{\rm + 0.09}}_{{\rm - 0.05}}$ \\ 
\hline

\end{tabular}
\caption{Schechter parameters of the MFs for the quiescent and
  elliptical galaxies between $z=0.2$ and $z=2$. The errors combined
  the one sigma Poissonian errors ($2\Delta ln\mathcal L=1$) as well
  as the uncertainties induced by the photo-z. Parameters $\alpha$
  listed without errors are set ``ad-hoc", and the errors on
  $log({\cal M}^*)$ and $\Phi^*$ are obtained by varying $\alpha$ by
  $\pm$0.5 around the fixed value. The morphological classification is
  obtained using the G-C parameters. However, we caution the
    reader that the errors are probably underestimated: the error
    budget is dominated by systematic effects (e.g. possible photo-z
    biases, systematic uncertainties related to the stellar mass
    estimate) that are not included here.
\label{schechterRed}}
\end{center}
\end{table*}

\begin{table*}[htb!]
\begin{center}
\begin{tabular}{l c c c c c c c} \hline

     &       &         &                       &         & \multicolumn{1}{c}{$log({\cal M}^*$}) &                  \multicolumn{1}{c}{$\Phi^*$} & $log(\rho^*)$  \\ 
type & z-bin &  Number & $log({\cal M}_{low})$ &  $\alpha$ &         (${\cal M}_\Sun$)             &             ($10^{-3} Mpc^{-3}$)        & (${\cal M}_\Sun Mpc^{-3}$)         \vspace{0.2cm} \\ \hline
intermediate&       0.2-0.4 &  5410 &   8.8 & -1.20$^{{\rm + 0.02}}_{{\rm - 0.02}}$ & 10.96$^{{\rm + 0.03}}_{{\rm - 0.03}}$ &  1.31$^{{\rm + 0.09}}_{{\rm - 0.09}}$ &  8.14$^{{\rm + 0.03}}_{{\rm - 0.03}}$ \\ 
activity&           0.4-0.6 &  4346 &   9.1 & -1.02$^{{\rm + 0.02}}_{{\rm - 0.02}}$ & 10.93$^{{\rm + 0.03}}_{{\rm - 0.03}}$ &  0.96$^{{\rm + 0.06}}_{{\rm - 0.06}}$ &  7.92$^{{\rm + 0.02}}_{{\rm - 0.03}}$ \\ 
     &              0.6-0.8 &  4837 &   9.3 & -0.90$^{{\rm + 0.03}}_{{\rm - 0.03}}$ & 10.85$^{{\rm + 0.02}}_{{\rm - 0.02}}$ &  1.02$^{{\rm + 0.06}}_{{\rm - 0.06}}$ &  7.84$^{{\rm + 0.02}}_{{\rm - 0.02}}$ \\ 
     &              0.8-1.0 &  5242 &   9.4 & -0.54$^{{\rm + 0.03}}_{{\rm - 0.03}}$ & 10.73$^{{\rm + 0.02}}_{{\rm - 0.02}}$ &  1.52$^{{\rm + 0.07}}_{{\rm - 0.07}}$ &  7.86$^{{\rm + 0.02}}_{{\rm - 0.02}}$ \\ 
     &              1.0-1.2 &  3826 &   9.5 & -0.44$^{{\rm + 0.04}}_{{\rm - 0.04}}$ & 10.77$^{{\rm + 0.02}}_{{\rm - 0.02}}$ &  1.05$^{{\rm + 0.04}}_{{\rm - 0.04}}$ &  7.74$^{{\rm + 0.02}}_{{\rm - 0.02}}$ \\ 
     &              1.2-1.5 &  4741 &   9.6 & -0.88$^{{\rm + 0.04}}_{{\rm - 0.04}}$ & 10.94$^{{\rm + 0.03}}_{{\rm - 0.03}}$ &  0.45$^{{\rm + 0.03}}_{{\rm - 0.03}}$ &  7.57$^{{\rm + 0.02}}_{{\rm - 0.02}}$ \\ 
     &              1.5-2.0 &  5019 &   9.8 & -1.03$^{{\rm + 0.04}}_{{\rm - 0.04}}$ & 11.02$^{{\rm + 0.03}}_{{\rm - 0.03}}$ &  0.23$^{{\rm + 0.02}}_{{\rm - 0.02}}$ &  7.40$^{{\rm + 0.02}}_{{\rm - 0.02}}$ \\ 
\hline
high&               0.2-0.4 &  2231 &   8.7 & -1.51$^{{\rm + 0.04}}_{{\rm - 0.04}}$ & 10.42$^{{\rm + 0.07}}_{{\rm - 0.07}}$ &  0.36$^{{\rm + 0.06}}_{{\rm - 0.06}}$ &  7.23$^{{\rm + 0.03}}_{{\rm - 0.04}}$ \\ 
activity&           0.4-0.6 &  4626 &   8.8 & -1.47$^{{\rm + 0.03}}_{{\rm - 0.03}}$ & 10.39$^{{\rm + 0.05}}_{{\rm - 0.05}}$ &  0.46$^{{\rm + 0.06}}_{{\rm - 0.06}}$ &  7.28$^{{\rm + 0.02}}_{{\rm - 0.02}}$ \\ 
     &              0.6-0.8 & 10261 &   8.9 & -1.48$^{{\rm + 0.02}}_{{\rm - 0.02}}$ & 10.49$^{{\rm + 0.03}}_{{\rm - 0.03}}$ &  0.65$^{{\rm + 0.05}}_{{\rm - 0.05}}$ &  7.53$^{{\rm + 0.01}}_{{\rm - 0.02}}$ \\ 
     &              0.8-1.0 & 12686 &   9.0 & -1.33$^{{\rm + 0.02}}_{{\rm - 0.02}}$ & 10.48$^{{\rm + 0.02}}_{{\rm - 0.02}}$ &  1.00$^{{\rm + 0.06}}_{{\rm - 0.06}}$ &  7.61$^{{\rm + 0.01}}_{{\rm - 0.01}}$ \\ 
     &              1.0-1.2 & 10335 &   9.2 & -1.29$^{{\rm + 0.02}}_{{\rm - 0.02}}$ & 10.48$^{{\rm + 0.02}}_{{\rm - 0.02}}$ &  0.93$^{{\rm + 0.06}}_{{\rm - 0.06}}$ &  7.56$^{{\rm + 0.01}}_{{\rm - 0.01}}$ \\ 
     &              1.2-1.5 & 14609 &   9.2 & -1.26$^{{\rm + 0.02}}_{{\rm - 0.02}}$ & 10.54$^{{\rm + 0.02}}_{{\rm - 0.02}}$ &  0.79$^{{\rm + 0.04}}_{{\rm - 0.04}}$ &  7.53$^{{\rm + 0.01}}_{{\rm - 0.01}}$ \\ 
     &              1.5-2.0 &  8697 &   9.8 & -1.30                            & 10.75$^{{\rm + 0.70}}_{{\rm - 0.3}}$ &  0.39$^{{\rm + 0.3}}_{{\rm - 0.3}}$ &  7.45$^{{\rm + 0.23}}_{{\rm - 0.16}}$ \\ 
\hline

\end{tabular}
\caption{Schechter parameters of star-forming galaxies (``intermediate
  activity'', ``high activity'') between $z=0.2$ and $z=2$. A
  parametrization of the total MF can be retrieved by summing the
  Schechter fit of the ``quiescent'' (given in Table
  \ref{schechterRed}), ``intermediate activity'' and ``high activity''
  galaxies. Errors are computed as in Table
    2. \label{schechterBlue}}
\end{center}
\end{table*}

\clearpage

\appendix

\section{Absolute magnitude estimate}\label{absmag}

At high redshift, the k-correction is one of the main sources of
systematic error in the absolute magnitude and rest-frame color
estimate. The k-correction depends on the galaxy SED, which is not
directly observed (Oke et Sandage 1968). In order to minimize the
uncertainty induced by the k-correction term, the rest-frame
luminosity at a given wavelength, $\lambda$, is derived from the
apparent magnitude observed at $\lambda \times (1+z)$ (appendix A of
Ilbert et al. 2005). With this procedure, the absolute magnitudes are
less dependent on the SED.  One drawback of this method is that the
uncertainty in the observed apparent magnitude is directly propagated
into the absolute magnitude. For this reason, the absolute magnitudes
were measured from one of the following bands $u^*g^+r^+i^+z^+K$ and
$3.6\mu m$ all of which have the highest signal-to-noise ratio and a
zero-point correction lower than 0.05 mag (Ilbert et al., 2009).

\section{Additional constraint in the template-fitting using the $24\mu m$ fluxes}\label{priorSFR}

The deep MIPS S-COSMOS data were taken during Spitzer Cycle 3 and
cover the full COSMOS 2-deg$^2$ (Aussel et al. 2009, in prep.). The
$24\mu m$ sources were detected with SExtractor (Bertin \& Arnouts
1996) and their fluxes measured with a PSF fitting technique (Le
Floc'h et al. 2009). The infrared luminosities $L_{\rm IR}^{\rm MIPS}$
were extrapolated from the 24 $\mu m$ fluxes using the Dale \& Helou
(2002) library (Le Floc'h et al. 2009) and converted into $SFR_{\rm
  IR}$ using the calibration from Kennicutt (1998).

\begin{figure}[htb!]
\includegraphics[width=13cm]{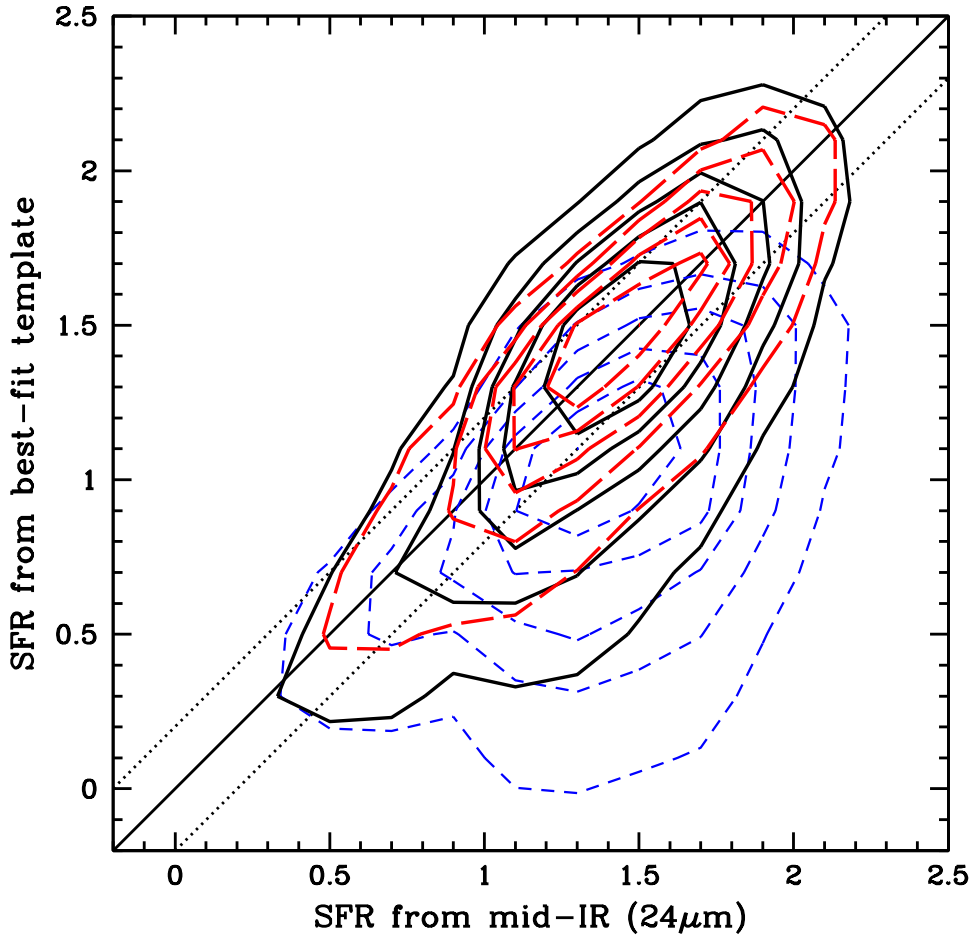}
\caption{SFR estimated from the best-fit
  template versus the SFR directly measured using the mid-infrared 24
  $\mu m$ MIPS flux. The short-dashed blue contours and solid black
  contours are obtained using the Charlot \& Fall (2000) extinction
  law and the Calzetti et al. (2000) extinction laws, respectively.
  The long-dashed red contours are obtained using the constraint on the
  $24\mu m$ flux. \label{LIRprior}}
\end{figure}

The stellar mass estimate and instantaneous SFR (hereafter
$SFR_{template}$) are derived from the best-fit template. We compared
$SFR_{template}$ and $SFR_{\rm IR}$ in order to decide which extinction
law to adopt -- either Calzetti et al. (2000) or Charlot \& Fall
(2000). Figure \ref{LIRprior} shows that the use of the Calzetti et al. (2000)
extinction law reduces the systematic offset between $SFR_{template}$
and $SFR_{\rm IR}$. Therefore, we favored the Calzetti et al. extinction
law for our analysis.

As a second step, we used the 24 $\mu m$ MIPS fluxes as an additional
constraint in the template fitting procedure. The goal of this
additional constraint is to remove possible degeneracies between ``old
and quiescent'' models and ``star-forming and dust extincted'' models.
For each template, we computed $L_{\rm IR}^{template}$, which is the
infrared luminosity which would be re-emitted in the infrared according to the
template, assuming that all of the UV light absorbed by dust is re-emitted
in the infrared and that the massive stars are the only source of infrared
emission. A likelihood, ${\cal L}$, is computed at each step of the
template fitting procedure. We multiplied this likelihood by the
probability to measure $L_{\rm IR}^{MIPS}$ for a given template:
\begin{equation}
{\cal L'} \propto {\cal L}(template, scaling, E(B-V)) \frac{exp(-\frac{(L_{\rm IR}^{MIPS}-L_{\rm IR}^{template})^2}{2 \times err^2})}{err \times \sqrt{2\pi}}
\end{equation}
where $err$ is the error on the $L_{\rm IR}^{MIPS}$ measurement (Le Floc'h
et al. 2009, in prep.). We added 0.2 dex in quadrature to the error on
$L_{\rm IR}^{MIPS}$ to take into account systematic uncertainties in
$L_{\rm IR}^{MIPS}$.  When the galaxy has no MIPS counterpart at more than
2$^{\prime\prime}$, an upper limit is applied to the $L_{\rm IR}^{template}$. This
upper-limit corresponds to the lowest $L_{\rm IR}^{MIPS}$ observable for a
sample selected at $F_{24\mu m}> 100\mu Jy$ (5$\sigma$ completeness
limit). No prior is applied if the optical counterpart is between
0.6$^{\prime\prime}$ and 2$^{\prime\prime}$ (less secure optical counterpart).

By construction, the comparison between $SFR_{template}$ and
$SFR_{\rm IR}$ is improved by this prior as shown in Figure
\ref{LIRprior}.

\section{The tool ALF}\label{ALF}

The selection of the galaxy sample at $F_{3.6\mu m}>1 \mu Jy$ defines
a limit in redshift above which the galaxies are too faint to be
observed. Statistical estimators are required to correct for the
incompleteness created by this flux limit.

We measured the luminosity and stellar mass functions using the tool
ALF ({\it Algorithm for Luminosity Function}) described in Ilbert et
al. (2005) and originally developed to measure the luminosity
functions from the VIMOS-VLT Deep Survey (le F\`evre et al. 2005). The
procedures used to compute the stellar MF or the luminosity function
are the same. The tool includes various estimators: the non-parametric
$1/V_{max}$, $C^+$, SWML and the parametric STY. The STY (Sandage et
al. 1979) and SWML (Efstathiou 1988) determine the MF by maximizing
the likelihood to observe a given stellar mass-redshift sample. The
STY estimator presupposes that the MF can be parametrized with a
Schechter function (Schechter 1976):
\begin{equation}
\Phi(M)dM=\Phi^*(M/M^*)^\alpha \; exp(-M/M^*) \; d(M/M^*).
\end{equation}
This parametrization allows us to describe the MF using three
parameters: $\alpha$ (slope), ${\cal M}^*$ (characteristic stellar
mass) and $\phi^*$ (normalization). The SWML is a non-parametric
estimate of the MF, useful to verify that a Schechter function is a
good representation of the data. The non-parametric 1/V$_{\rm max}$
estimator (Schmidt 1968) is the most widely used because of its
simplicity. The 1/V$_{\rm max}$ is the inverse sum of the volume in
which each galaxy could be observed. The 1/V$_{\rm max}$ is the only
estimator directly normalized. Lynden-Bell (1971) derived the
non-parametric C$^-$ method to overcome the assumption of a uniform
galaxy distribution derived using 1/V$_{\rm max}$ (we used a slightly
modified version called C$^+$, Zucca et 1997). The implementation of
these estimators is detailed in appendix A of Ilbert et al. (2005).

\begin{figure}[htb!]
\begin{tabular}{c c}
\includegraphics[width=7.9cm]{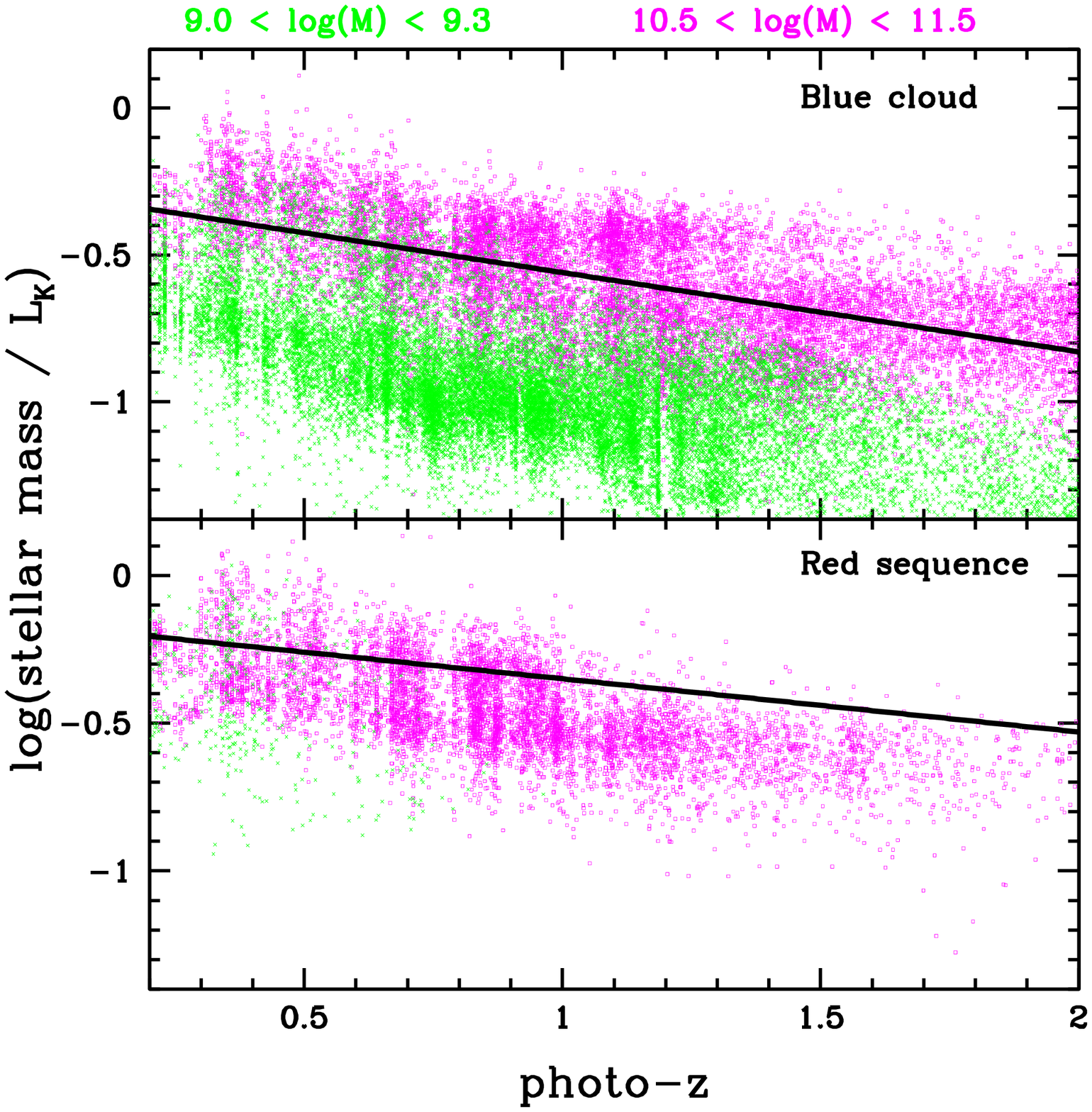}
\includegraphics[width=7.9cm]{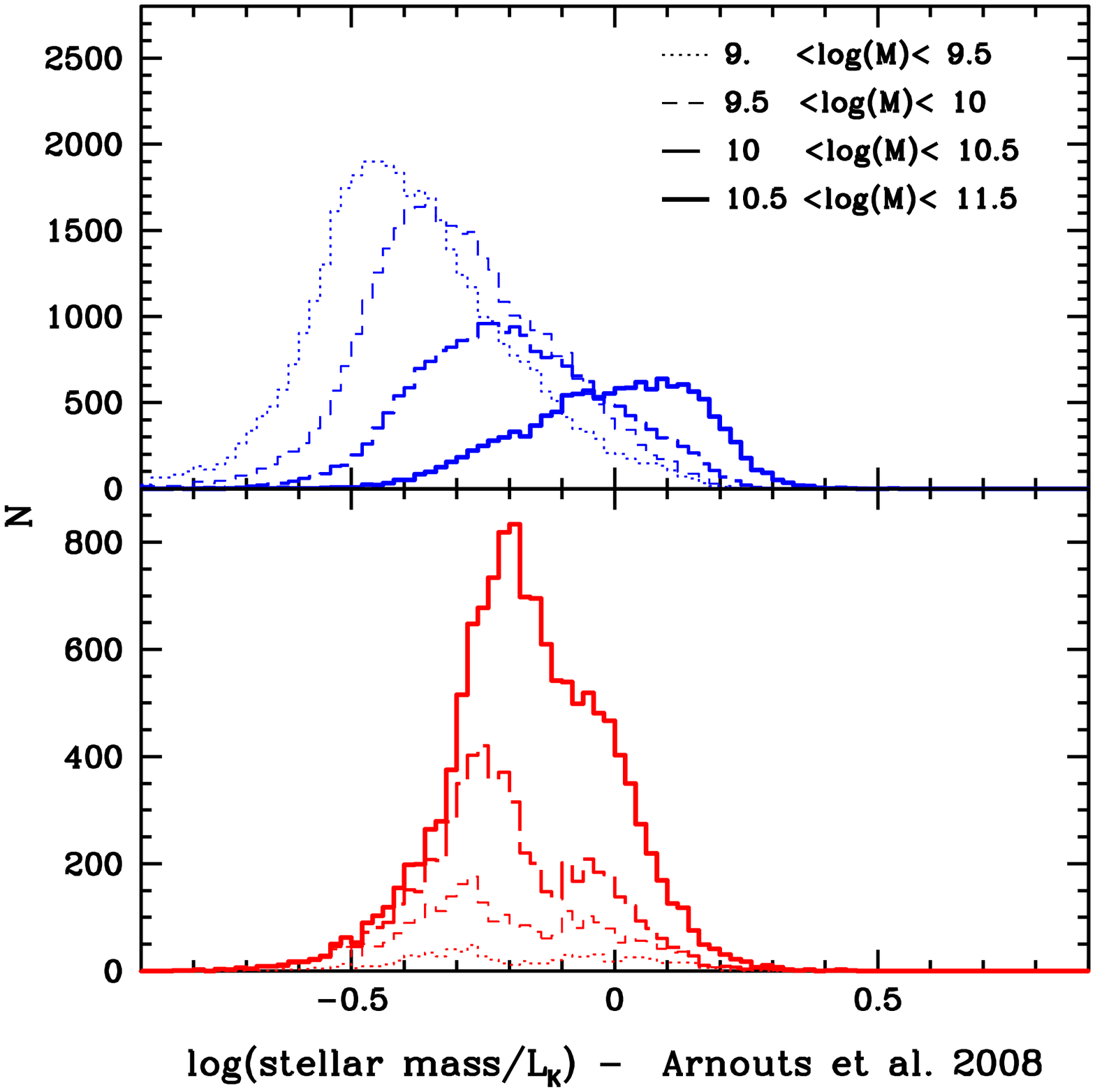}
\end{tabular}
\caption{{\bf Left panel}: mass-to-light ratio as a function of
  redshift. The stellar masses are computed from a SED fitting
  procedure. The luminosity is measured in the rest-frame $K_{\rm
    s}$-band. The top and bottom panels correspond to the blue cloud
  and red sequence galaxies, respectively. The solid lines correspond
  to the empirical relations derived by Arnouts et al. (2007). The
  relation of Arnouts et al. (2007) is derived for the massive
  galaxies brighter than $m_{3.6 \mu m}<21.5$. The magenta crosses and
  the green circles are the galaxies selected in stellar mass ranges
  $9.0<log({\cal M})<9.5$ and $10.5<log({\cal M})<11.5$,
  respectively. {\bf Right panel}: difference between the observed
  mass-to-light ratio and the analytical relation established by
  Arnouts et al. (2007). The top and bottom panels correspond to the
  star-forming and quiescent galaxies, respectively. The difference is
  shown per bin of stellar mass.\label{ML_z}}
\end{figure}

\section{Mass-to-light ratio}\label{masstolight}

The $K$-band luminosity is not a direct tracer of the stellar
mass. Numerous analyzes have used an analytical parametrization of the
mass-to-light ratio in order to derive the stellar masses
(e.g. Kochanek et al. 2001, Driver et al. 2007, Arnouts et al. 2007).

The left panel of Figure \ref{ML_z} shows the mass-to-light ratio as a
function of redshift (the stellar masses are derived as described in
section.\ref{stellarmass}). We observed a decrease of the
mass-to-light ratio with redshift, in good agreement with the
empirical relations derived by Arnouts et al. (2007). We also found
that the mass-to-light ratio increases with the stellar mass. This
trend is expected since the high mass galaxies are older and have  a larger
mass-to-light ratio. The right panel of Figure \ref{ML_z} shows the
difference between the observed mass-to-light ratio and the analytical
relation derived by Arnouts et al. (2007). For the quiescent galaxies,
the median difference is below 0.1 dex. The median difference is also
below 0.1 dex for the star-forming galaxies ($log({\cal
M})>10.5$), but the median difference is shifted below -0.3 dex at low
masses $log({\cal M})<9.5$, showing that this relation is not anymore
a good proxy for the low mass galaxies.

We compared our stellar mass densities for star-forming galaxies with
results from the literature in Figure \ref{LD_Litt}. The stellar mass
densities of Arnouts et al. (2007) are systematically 0.2 dex higher
than our measurements. Arnouts et al. (2007) based their measurements
on the $K$-band LF converted into stellar mass density using the
mass-to-light relation discussed above. At low masses ($log({\cal
M})<10$), we found mass-to-light ratio lower by 0.2-0.4 dex than those
estimated from the Arnouts et al. (2007) relation.  Since star-forming
galaxies are mostly low/intermediate mass galaxies (steep slope of
their MF), the dependency of the mass-to-light ratio on the mass
likely explains this offset of 0.2 dex in the stellar mass densities.

\section{The quiescent MFs in the GOODS and COSMOS fields}\label{GOODS}

Two possible sources of incompleteness could affect our estimate
  of the quiescent MFs at $z>1$: the depth of the optical catalogue
  and the confusion in the IRAC images. We checked the validity of our
  selection criteria using the GOODS datasets (Giavalisco et
  al. 2004). We used the public catalogues FIREWORKS (Wuyts et
  al. 2008) and MUSIC (Santini et al. 2009). The IRAC photometry in
  these catalogues is less affected by the confusion since specific
  softwares allow an accurate ``PSF-matching'' (e.g. convphot in
  Santini et al. 2009). The GOODS optical data are also deeper than
  COSMOS (90\% complete in $ACS/F775W$ at 26.5 mag).

We selected the FIREWORKS and MUSIC catalogues at $F(3.6\mu m)>1\mu
Jy$, which is the selection limit of our study. We computed the
photo-z using the code "Le Phare'' and the same setting (templates,
extinction, ...) as Ilbert et al. (2009). The rest-frame colors and
stellar masses were estimated following exactly the method described
in section \ref{method}. We also classified galaxies according to
their dust-corrected $NUV-R$ rest-frame colors as in section
\ref{template}.

\begin{figure}[ht] 
\includegraphics[height=16cm]{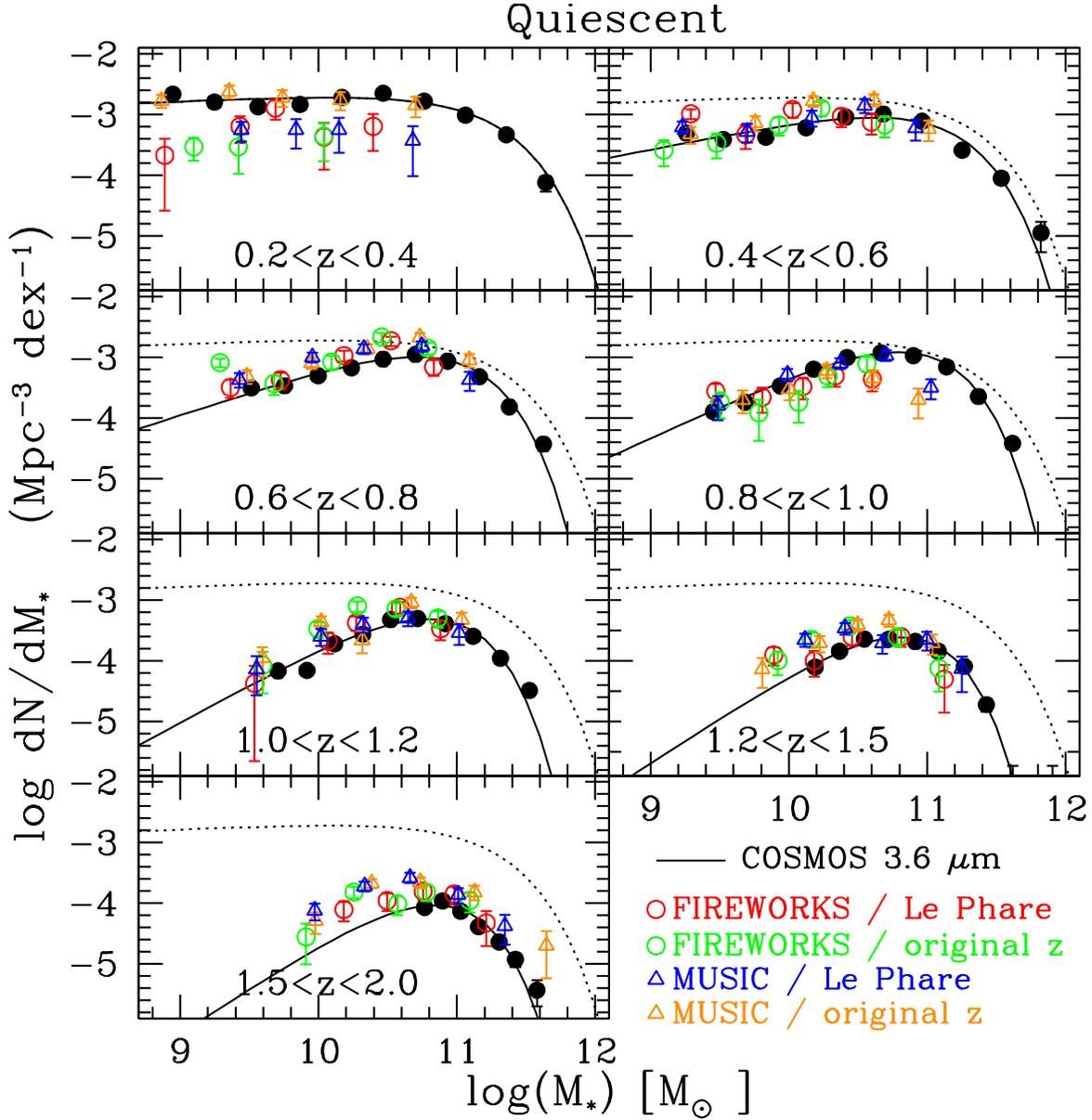}
\caption{Stellar mass functions of the quiescent population in
    the GOODS and COSMOS fields. The black filled circles correspond
    to the S-COSMOS estimate of the quiescent MFs. The open points are
    the quiescent MFs measured in the GOODS field. Red and green
    circles are an estimate based on the FIREWORKS catalogue using
    ``Le Phare'' and the Wuyts et al. (2008) photo-z,
    respectively. Blue and orange triangles are an estimate based on
    the MUSIC catalogue using ``Le Phare'' and the Santini et
    al. (2009) photo-z, respectively.}
\label{MFGOODS}
\end{figure}

First, we find that the contribution of galaxies fainter than $I=26.5$
is negligible at the considered stellar mass limits (see table
\ref{schechterRed}). Then, we derive the quiescent mass functions in
GOODS (using both MUSIC and FIREWORKS catalogues, as well as their
original photo-z). The comparison with the S-COSMOS MFs is shown in
Figure \ref{MFGOODS}. The results are in excellent agreement. The
differences could be easily explained by cosmic variance since the
GOODS field covers only 160 arcmin$^2$. The only significant
difference is seen using the GOODS-MUSIC catalogue and their own set
of photo-z at $z>1.2$ (a factor $\sim 1.5$ in normalisation). From
this comparison, we conclude that our results don't suffer from any
significant incompleteness at $z>1$ and that the rapid assembly of the
quiescent population at $1<z<2$ can't be explained by incompleteness
or confusion effects.

\clearpage

\end{document}